\documentclass[useAMS,usenatbib]{mn2e}

\usepackage[normalem]{ulem} 

\usepackage{graphicx}
\usepackage{hyperref}
\usepackage{color,epsfig}
\usepackage[utf8]{inputenc}
\usepackage{verbatim}
\usepackage{lmodern}
\usepackage{amssymb}

\newcommand{\be}{\begin{equation}}
\newcommand{\ee}{\end{equation}}

\newcommand*\diff{\mathop{}\!\mathrm{d}}

\newcommand*{\eqref}[1]{(\ref{#1})}

\newcommand{\vect}[1]{\mathbf{#1}}

\def\msun{{\,{\rm M}_\odot}}
\def\rsun{{\,{\rm R}_\odot}}
\def\gcm3{\, \rm g \, cm^{-3}}

\def\rt{R_{\rm t}}
\def\rg{R_{\rm g}}
\def\rp{R_{\rm p}}

\def\mh{M_{\rm h}}

\def\mstar{M_{\star}}
\def\rstar{R_{\star}}

\def\tmin{t_{\rm min}}

\def\ledd{L_{\rm Edd}}

\def\ergs{\rm erg \, s^{-1}}

\def\msun{M_{\odot}}

\def\jacc{j_{\rm acc}}

\def\vaovc{v_{\rm A}/v_{\rm c}}

\def\tev{t_{\rm ev}}

\title[Long-term stream evolution in TDEs]{Long-term stream evolution in tidal disruption events}
\author[Clément Bonnerot, Elena M. Rossi and Giuseppe Lodato]{Clément Bonnerot$^{1}$\thanks{E-mail: bonnerot@strw.leidenuniv.nl}, Elena M. Rossi$^{1}$ and Giuseppe Lodato$^{2}$\\
$^{1}$Leiden Observatory, Leiden University, PO Box 9513, 2300 RA, Leiden, the Netherlands\\
$^{2}$Dipartimento
 di Fisica, Università Degli Studi di Milano, Via Celoria, 16, Milano, 20133, Italy
}
\begin{document}

\date{ Accepted ?. Received ?; in original form ?}

\pagerange{\pageref{firstpage}--\pageref{lastpage}} \pubyear{2014}

\maketitle

\label{firstpage}

\begin{abstract}

\noindent A large number of tidal disruption event (TDE) candidates have been observed recently, often differing in their observational features. Two classes appear to stand out: X-ray and optical TDEs, the latter featuring lower effective temperatures and luminosities. These differences can be explained if the radiation detected from the two categories of events originates from different locations. In practice, this location is set by the evolution of the debris stream around the black hole and by the energy dissipation associated with it. In this paper, we build an analytical model for the stream evolution, whose dynamics is determined by both magnetic stresses and shocks. Without magnetic stresses, the stream always circularizes. The ratio of the circularization timescale to the initial stream period is $t_{\rm ev}/t_{\rm min} = 8.3 (M_{\rm h}/10^6 M_{\odot})^{-5/3} \beta^{-3}$, where $M_{\rm h}$ is the black hole mass and $\beta$ is the penetration factor. If magnetic stresses are strong, they can lead to the stream ballistic accretion. The boundary between circularization and ballistic accretion corresponds to a critical magnetic stresses efficiency $v_{\rm A}/v_{\rm c} \approx 10^{-1}$, largely independent of $M_{\rm h}$ and $\beta$. However, the main effect of magnetic stresses is to accelerate the stream evolution by strengthening self-crossing shocks. Ballistic accretion therefore necessarily occurs on the stream dynamical timescale. The shock luminosity associated to energy dissipation is sub-Eddington but decays as $t^{-5/3}$ only for a slow stream evolution. Finally, we find that the stream thickness rapidly increases if the stream is unable to cool completely efficiently. A likely outcome is its fast evolution into a thick torus, or even an envelope completely surrounding the black hole.
\end{abstract}


\section{Introduction}

Two-body encounters between stars surrounding a supermassive black hole occasionally result in one of these stars being scattered on a plunging orbit towards the central object. If this star is brought too close to the black hole, the strong tidal forces exceed its self-gravity force, leading to the star's disruption. About half of the stellar material ends up being expelled. The remaining fraction stays bound and returns the black hole as an extended stream of gas \citep{rees1988} with a mass fallback rate decaying as $t^{-5/3}$. This bound material is expected to be accreted, resulting in a luminous flare. Such tidal disruption events (TDEs) contain information on the black hole and stellar properties. While white dwarf tidal disruptions necessarily involve black holes with low masses $\mh \lesssim 10^5 \msun$ \citep{macleod2015}, TDEs involving giant stars are best suited to probe the higher end of the black hole mass function, with $\mh \gtrsim 10^8 \msun$ \citep{macleod2012}. However, the latter might be averted by the dissolution of the debris into the background gaseous environment through Kelvin-Helmholtz instability, likely dimming the associated flare \citep{bonnerot2016_2}. TDEs also represent a unique probe of accretion and relativistic jets physics. Additionally, they could provide insight into bulge-scale stellar processes through the rate at which stars are injected into the tidal sphere to be disrupted.\\

The number of candidate TDEs is rapidly growing \citetext{see \citealt{komossa2015} for a recent review}. Most of the detected electromagnetic signals peak in the soft X-ray band \citep{komossa1999,cappelluti2009,esquej2008,maksym2010,saxton2012} and at optical and UV wavelengths \citep{gezari2006,gezari2012,van_velzen2011,cenko2012,arcavi2014,holoien2016}. In addition, a small fraction of candidates shows both optical and X-ray emission \citetext{e.g. ASSASN-14li, \citealt{holoien2016}}. Finally, TDEs have been detected in the hard X-ray to $\gamma$-ray band \mbox{\citep{cenko2012b,bloom2011}}.\\

The classical picture for the emission mechanism relies on an efficient circularization of the bound debris as it falls back to the disruption site \citep{rees1988,phinney1989}. In this scenario, the emitted signal comes from an accretion disc that forms rapidly from the debris at $\sim 2 \rp$, where $\rp$ denotes the pericentre of the initial stellar orbit. The  argument for rapid disc formation involves self-collision of the stream debris due to relativistic precession at pericentre. This picture is able to explain the observed properties of X-ray TDE candidates, which feature an effective temperature $T_{\rm eff} \approx 10^5 \, \rm K$ with a luminosity up to $L \approx 10^{44} \, \ergs$. However, it is inconsistent with the emission detected from optical TDEs, with $T_{\rm eff} \approx 10^4\, \rm  K$ and $L \approx 10^{43} \, \ergs$. This is because the disc emits mostly in the X-ray, with a small fraction of the radiation escaping as optical light, typically only $\lesssim 10^{41} \, \ergs$ in terms of luminosity \citep[their figure 2]{lodato2011}.\\

The puzzling features of optical TDEs have motivated numerous investigations. Several works argue that optical photons are emitted from a shell of gas surrounding the black hole at a distance $\sim 100 \, \rp$. This envelope could reprocess the X-ray emission produced by the accretion disc, giving rise to the optical signal. This reprocessing layer is a natural consequence of several mechanisms, such as winds launched from the outer parts of the accretion disc \citep{strubbe2009,lodato2011,miller2015} and the formation of a quasistatic envelope from the debris reaching the vicinity of the black hole \citep{loeb1997,guillochon2014,coughlin2014,metzger2015}. As noticed by \citet{metzger2015}, the latter possibility is motivated by recent numerical simulations that find that matter can be expelled at large distances from the black hole during the circularization process \citep{ramirez-ruiz2009,bonnerot2016_1,hayasaki2015,shiokawa2015,sadowski2015}\\

Another interesting idea has been put forward by \citet{piran2015}, although it has been proposed for the first time by \citet{lodato2012}. They argue that the optical emission could come from energy dissipation associated with the circularization process, and be produced by shocks occurring at distances much larger than $\rp$. Furthermore, since the associated luminosity relates to the debris fallback rate, they argue that it should scale as $t^{-5/3}$ as found observationally \citetext{e.g. \citealt{arcavi2014}}. Such outer shocks are expected for low apsidal precession angles, which was shown to be true as long as the star only grazes the tidal sphere \citep{dai2015}. Owing to the weakness of such shocks, these authors suggest that the debris could retain a large eccentricity for a significant number of orbits. Recently, this picture was claimed to be consistent with the X-ray and optical emission detected from ASSASN-14li \citep{krolik2016}. Nevertheless, the absence of X-ray emission in most optical TDEs is hard to reconcile with this picture, since viscous accretion should eventually occur, leading to the emission of X-ray photons. For this reason, \citet{svirski2015} proposed that magnetic stresses are able to remove enough angular momentum from the debris to cause its ballistic accretion with no significant emission in tens of orbital times. In their work, energy loss via shocks has been omitted. However, they are likely to occur as the stream self-crosses due to relativistic precession. This provides an efficient circularization mechanism that could give rise to X-ray emission. This is all the more true that the pericentre distance decreases as magnetic stresses act on the stream, thus strengthening apsidal precession and the resulting shocks.\\

In this paper, we present an analytical treatment of the long-term evolution of the steam of debris under the influence of both shocks and magnetic stresses. We show that, even if the stream retains a significant eccentricity after the first self-crossing, subsequent shocks are likely to further shrink the orbit. Furthermore, the main impact of magnetic stresses is found to be the acceleration of the stream evolution via a strengthening of self-crossing shocks. If efficient enough, magnetic stresses can also lead to ballistic accretion. However, this necessarily happens in the very early stages of the stream evolution. In addition, we demonstrate that a $t^{-5/3}$ decay of the shock luminosity light curve is favoured for a slow stream evolution, favoured for grazing encounters with black hole masses $\lesssim 10^6 \msun$. This decay law is in general hard to reconcile with ballistic accretion that occurs on shorter timescales. Finally, we demonstrate that if the excess thermal energy injected by shocks is not efficiently radiated away, the stream rapidly thickens to eventually form a thick structure.

This paper is organised as follows. In Section \ref{model}, the stream evolution model under the influence of shocks and magnetic stresses is presented. In Section \ref{results}, we investigate the influence of the different parameters on the stream evolution and derive the observational consequences. In addition, we investigate the influence of inefficient cooling on the stream geometry. Finally, Section \ref{conclusion} contains the discussion of these results and our concluding remarks.

\section{Stream evolution model}
\label{model}

A star is disrupted by a black hole if its orbit crosses the tidal radius $\rt=\rstar(\mh/\mstar)^{1/3}$, where $\mh$ denotes the black hole mass, $\mstar$ and $\rstar$ being the stellar mass and radius. Its pericentre can therefore be written as $\rp=\rt/\beta$, where $\beta>1$ is the penetration factor. During the encounter, the stellar elements experience a spread in orbital energy $\Delta \epsilon = G \mh R_{\star} / R^2_t$, given by their depth within the black hole potential well at the moment of disruption \citep{lodato2009,stone2013}. The debris therefore evolves to form an eccentric stream of gas, half of which falls back towards to black hole.

The most bound debris has an energy $-\Delta \epsilon$. It reaches the black hole after $\tmin=2 \pi G \mh (2 \Delta \epsilon)^{-3/2}$ from the time of disruption, following Kepler's third law. Due to relativistic apsidal precession, it then continues its revolution around the black hole on a precessed orbit. This results in a collision with the part of the stream still infalling. This first self-crossing leads to shocks that dissipate part of the debris orbital energy into heat. The resulting stream moves closer to the black hole, its precise trajectory depending on the amount of orbital energy removed. As the stream continues to orbit the black hole, more self-crossing shocks must happen due to apsidal precession at each pericentre passage.

\begin{figure}
\epsfig{width=0.47\textwidth, file=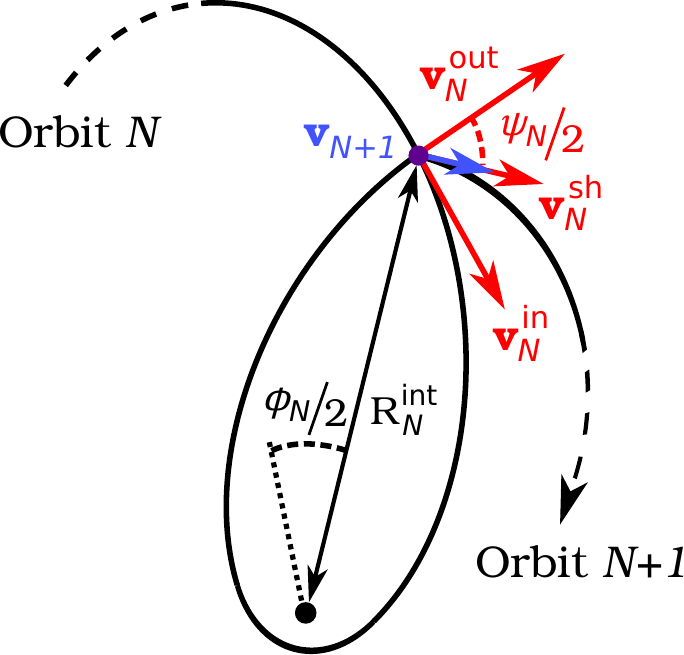}
\caption{Sketch illustrating the stream evolution model as a succession of orbits. Orbit $N+1$ follows orbit $N$ after an energy loss through shocks and an angular momentum loss through magnetic stresses. The associated velocity changes are depicted in red and blue respectively. While the stream is on orbit $N$, it precesses by an angle $\phi_N$, given by equation \eqref{precang}. As a result, the stream self-crosses at the intersection point, indicated by the purple point. It occurs at a distance $R^{\rm int}_N$ from the black hole, given by equation \eqref{rint}, and with a collision angle $\psi_N$. The post-shock velocity $\vect{v}^{\rm sh}_N$ is obtained  from the velocity $\vect{v}^{\rm in}_N$ and $\vect{v}^{\rm out}_N$ of the two colliding components according to equation \eqref{vsh}. Immediately after, the stream undergoes magnetic stresses that reduce this velocity to $\vect{v}_{N+1}$, given by equation \eqref{vnew} and defining the initial velocity of orbit $N+1$.}
\label{fig1}
\end{figure}

We therefore model the evolution of the stream as a succession of Keplerian orbits, starting from that of the most bound debris. From one orbit to the next, the stream orbital parameters change according to both shocks and magnetic stresses, as described in Sections \ref{shocks} and \ref{stresses} respectively. This is illustrated in Fig. \ref{fig1} that shows two successive orbits of the stream, labelled $N$ and $N+1$. Knowing the orbital changes between successive orbits allows to compute by iterations the orbital parameters of any orbit $N$. This iteration is performed until the stream reaches its final outcome, defined by the stopping conditions presented in Section \ref{outcome}. In the following, variables corresponding to orbit $N$ are indicated by the subscript ``$N$''.

The initial orbit, corresponding to $N=0$, is that of the most bound debris. It has a pericentre $R^{\rm p}_0$ equal to that of the star $\rp$ and an eccentricity $e_0 = 1 - (2/\beta)(\mh/\mstar)^{-1/3}$. Its energy is
\be
\epsilon_0=-\Delta \epsilon \propto M^{1/3}_{\rm h},
\label{en0}
\ee
while, using $e_0 \approx 1$, its angular momentum can be approximated as
\be
j_0 \approx \sqrt{2 G M_{\rm h} \rp} \propto M^{2/3}_{\rm h} \beta^{-1/2}.
\label{j0}
\ee
From this initial orbit, the orbital parameters of any orbit $N$ are computed iteratively. Its energy and angular momentum are given by
\be
\epsilon_N=-\frac{G M_{\rm h}}{2 a_N},
\label{energy}
\ee
\be
j_N=\sqrt{G M_{\rm h}a_N(1-e^2_N)},
\label{angmom}
\ee
respectively as a function of the semi-major axis $a_N$ and eccentricity $e_N$ of the stream. The apocentre and pericentre distances of orbit $N$ are by definition $R^{\rm a}_N= a_N (1+e_N)$ and $R^{\rm p}_N = a_N (1-e_N)$ respectively. At these locations, the stream has velocities $v^{\rm a}_N= (G \mh /a_N)^{1/2} ((1-e_N)/(1+e_N))^{1/2}$ and $v^{\rm p}_N= (G \mh /a_N)^{1/2} ((1+e_N)/(1-e_N))^{1/2}$.

Our assumption of a thin stream moving on Keplerian trajectories requires that pressure forces are negligible compared to gravity. This is legitimate as long as the excess thermal energy produced by shocks is radiated efficiently away from the gas. The validity of this approximation is the subject of Section \ref{cooling}.

Moreover, our treatment of the stream evolution neglects the dynamical impact of the tail of debris that keeps falling back long after the first collision. This fact will be checked a posteriori in Section \ref{observational}. It can already be justified qualitatively here through the following argument. At the moment of the first shock, the tail and stream densities are similar. However, later in the evolution, the tail gets stretched resulting in a density decrease. On the other hand, the stream gains mass and moves closer to the black hole as it loses energy. As a consequence, its density increases. The tail therefore becomes rapidly much less dense than the stream, which allows to neglect its dynamical influence on the stream evolution.

\subsection{Shocks}
\label{shocks}

Apsidal precession causes the first self-crossing shock that makes the debris produced by the disruption more bound to the black hole. The subsequent evolution of the stream is affected by a similar process. When an element of the stream passes at pericentre, its orbit precesses, causing its collision with the part of the stream still moving towards pericentre. Once all the stream matter has passed through this intersection point, it continues on a new orbit.

Suppose that the stream is on orbit $N$. To determine the change of orbital parameters due to shocks as the stream self-crosses, we use a treatment similar to that used by \citet{dai2015} to predict the orbit resulting from the first self-intersection. This method is also inspired from an earlier work by \citet{kochanek1994}. It is illustrated in Fig. \ref{fig1}, with the associated change in velocity shown in red. As orbit $N$ precesses by an angle\footnote{This expression is derived in the small-angle approximation. This is legitimate since our study
mainly focuses on outer shocks, for which $\phi_N$ remains smaller than about 10 degrees. In the case of strong shocks, the stream evolution is fast independently on the precise value of $\phi_N$.}\citep[p. 232]{hobson2006} 
\be
\phi_N=\frac{6 \pi G M_{\rm h}}{a_N(1-e^2_N)c^2},
\label{precang}
\ee
it intersects the remaining part of the stream at a distance from the black hole
\be
R^{\rm int}_N=\frac{a_N(1-e^2_N)}{1-e_N \cos(\phi_N/2)}.
\label{rint}
\ee
The angle in the denominator is computed from a reference direction that connects the pericentre and apocentre of orbit $N$. At this point, the infalling and outflowing parts of the stream collide. Following \citet{dai2015}, we assume this collision to be completely inelastic. Momentum conservation then sets the resulting velocity to
\be
\vect{v}^{\rm sh}_{N}=\frac{\vect{v}^{\rm in}_N+\vect{v}^{\rm out}_N}{2},
\label{vsh}
\ee
where $\vect{v}^{\rm in}_N$ and $\vect{v}^{\rm out}_N$ denote the velocity of the inflowing and outflowing components respectively. Equation \eqref{vsh} assumes that the two components have equal masses. This is justified since they are part of the same stream. Although this stream might be inhomogeneous shortly after the first shock, inhomogeneities are likely to be suppressed later in its evolution. Note that conservation of momentum implies conservation of angular momentum since this velocity change occurs at a fixed position.

According to equation \eqref{vsh}, the post-shock velocity is given by $|\vect{v}^{\rm sh}_N|= |\vect{v}_N| \cos(\psi_N/2)$, where $\psi_N$ is the collision angle between $\vect{v}^{\rm in}_N$ and $\vect{v}^{\rm out}_N$ and $|\vect{v}_N|$ denotes the velocity at the intersection point, equal to $|\vect{v}^{\rm in}_N|$ and $|\vect{v}^{\rm out}_N|$ because of energy conservation along the Keplerian orbit. Therefore, the energy removed from the stream during the collision is
\be
\Delta \epsilon_N = \frac{1}{2} \vect{v}^2_N \sin^2(\psi_N/2).
\label{deltaen_vn}
\ee
Using $\epsilon_N = \vect{v}^2_N/2 - G \mh /R^{\rm int}_N$ and equation \eqref{rint} combined with $\sin^2(\psi_N/2)=e^2_N \sin^2(\phi_N/2)/(1+e^2_N-2e_N\cos(\phi_N/2))$, this can be rewritten as
\be
\Delta \epsilon_N = \frac{e^2_N}{2} \left( \frac{G \mh}{j_N} \right)^2  \sin^2(\phi_N/2),
\label{deltaen}
\ee
which also makes use of the relation $(G \mh)^2 (1-e^2_N) = -2 j^2_N \epsilon_N$. Equation \eqref{deltaen} has the advantage of depending only on the orbital parameters of orbit $N$. It will be used in Section \ref{eqdif} to find an equivalent differential equation describing the stream evolution. In addition, equation \eqref{deltaen} implies that $\Delta \epsilon_N$ is largely independent of $N$ when the stream angular momentum is unchanged, which is the case if magnetic stresses do not affect its evolution. This is because $e^2_N$ varies only weakly with $N$ while $\phi_N$ only depends on $j_N$ as can be seen by combining equations \eqref{angmom} and \eqref{precang}. The constant value of $\Delta \epsilon_N$ can then be obtained by evaluating equation \eqref{deltaen} at $N=0$. Simplifying by the small angle approximation $\sin \theta \approx \theta$, it is given by
\be
\Delta \epsilon_0 = \left( \frac{9 \pi^2}{16 c^4}\right) e^2_0 \left(\frac{G \mh}{\rp} \right)^3 \propto \mh^2 \beta^3,
\label{deltaenzero}
\ee
using equation \eqref{j0} and $e_0 \approx 1$. The fact that $\Delta \epsilon_N \approx \Delta \epsilon_0$ will be used in Section \ref{dynamical} to find an analytical expression for the circularization timescale of the stream in the absence of magnetic stresses.

\subsection{Magnetic stresses}
\label{stresses}

Magnetic stresses act on the stream, leading to angular momentum transport outwards. To evaluate the orbital change induced by this mechanism, we follow \citet{svirski2015}. Consider a stream section covering an azimuthal angle $\delta \phi$ and located at a distance $R=j^2/(G \mh) / (1+e \cos(\theta))$ from the black hole, $j$ denoting its specific angular momentum and $\theta$ its true anomaly. This section loses specific angular momentum at a rate
\be
\diff j/\diff t= (\diff \mathcal{G}/\diff R) /(\Sigma R \delta \phi).
\label{djdtini}
\ee
In this expression, $\mathcal{G}$ is the rate of angular momentum transport outwards, given by
\be
\mathcal{G}=\int_{-\Delta z}^{\Delta z} R \mathcal{M}_{\vect{\hat{n}} \vect{\hat{t}}} |\vect{\hat{r}} \times \vect{\hat{t}}| R \delta \phi \diff z,
\label{grate}
\ee
where $\Delta z$ denotes the vertical extent of the stream, $\vect{\hat{n}}$ and $\vect{\hat{t}}$ are unit vectors normal and tangential to the stream section considered while $\vect{\hat{r}}$ is in the radial direction. $\mathcal{M}_{\vect{\hat{n}} \vect{\hat{t}}} = -B_\vect{\hat{n}} B_\vect{\hat{t}} / (4\pi)$ denotes the $\vect{\hat{n}}$-$\vect{\hat{t}}$ component of the Maxwell tensor, $B_\vect{\hat{n}}$ and $B_\vect{\hat{t}}$ being the normal and tangential component of the magnetic field. The term $|\vect{\hat{r}} \times \vect{\hat{t}}|=(1+e \cos \theta)/(1+e^2+2e \cos \theta)^{1/2}$ is required since only the component of $\vect{\hat{t}}$-momentum orthogonal to $\vect{\hat{r}}$ contributes to the angular momentum. Combining equations \eqref{djdtini} and \eqref{grate} then leads to
\be
\frac{\diff j}{\diff t}= \alpha_{\rm mag} |\vect{\hat{r}} \times \vect{\hat{t}}| v^2_{\rm A} ,
\label{djdt}
\ee
where $\alpha_{\rm mag}=-2 B_\vect{\hat{n}} B_\vect{\hat{t}}/B^2$ and $v^2_{\rm A}=\int_{-\Delta z}^{\Delta z} B^2 \diff z/(4 \pi \Sigma)$ is the squared Alfvén velocity. The angular momentum $\Delta j$ lost by the stream section\footnote{More precisely, angular momentum is transferred outwards from the bulk of the stream to a gas parcel of negligible mass. It is therefore a fair assumption to assume that this angular momentum is lost.} in one period is then obtained by integrating equation \eqref{djdt}. Using the chain rule to combine equation \eqref{djdt} with Kepler's second law $\diff \theta/\diff t=j/R^2$ and integrating over $\theta$, it is found to be
\be
\Delta j=K_e  \left(\frac{v_{\rm A}}{v_{\rm c}}\right)^2 j ,
\label{deltaj}
\ee
where
\be
K_e\equiv\alpha_{\rm mag} \int_0^{2\pi} f_e(\theta)  \diff \theta,
\label{ke}
\ee
with $f_e(\theta)\equiv(1+e^2+2e \cos \theta)^{-1/2}$ and $v_{\rm c}=(G\mh/R)^{1/2}$ being the circular velocity at $R$. Equation \eqref{deltaj} has been obtained by assuming that $\alpha_{\rm mag}$ and $v_{\rm A}/v_{\rm c}$ are independent of $R$. In the following, we set $\alpha_{\rm mag}=0.4$, as motivated by magnetohydrodynamical simulations \citep{hawley2011}. The value of $v_{\rm A}/v_{\rm c}$ is varied from $10^{-2}$ to 1. This range of values relies on the assumption that magneto-rotational instability has fully developed at its fastest rate associated to its most disruptive mode. The former is reached for $v_{\rm A} k \simeq v_{\rm c}/R$, $k$ being the wavenumber of the instability \citep{balbus1998}. The latter corresponds to the lowest wavenumber available, that is $k \simeq 1/H$ where $H$ denotes the width of the stream. Therefore, $v_{\rm A}/v_{\rm c} \simeq H/R$, which likely varies from $10^{-2}$ to 1. It is however possible that the MRI did not have time to reach saturation in the early stream evolution since it requires about 3 dynamical times \citep{stone1996}. This would lead to lower values of $v_{\rm A}/v_{\rm c}$. Since $f_e(\pi)/f_e(0)=(1+e)/(1-e)\gg1$ for $1-e\ll1$, the integrand in equation \eqref{ke} is the largest for $\theta \approx \pi$. As noticed by \citet{svirski2015}, this means that the angular momentum loss happens mostly close to apocentre as long as the eccentricity is large, which is true during the stream evolution. Instead, if the stream reaches a nearly circular orbit, angular momentum is lost roughly uniformly along the orbit. This argument will be used in Section \ref{outcome} to define one of the stopping criterion of the iteration.

Since magnetic stresses act mostly at apocentre, we implement it as an instantaneous angular momentum loss at this location. The angular momentum removed from orbit $N$ is then obtained from equation \eqref{deltaj} by $\Delta j_N = K_e (v_{\rm A}/v_{\rm c})^2 j_N$. The post-shock velocity given by equation \eqref{vsh} has no radial component, which can also be seen from Fig. \ref{fig1}. This implies that the apocentre of each orbit is located at the self-crossing point. Angular momentum loss therefore amounts to reducing the post-shock velocity given by equation \eqref{vsh} by a factor $1-\Delta j_N / j_N$. This defines the initial velocity of orbit $N+1$
\be
\vect{v}_{N+1}= \max \left(0,1-\frac{\Delta j_N}{j_N}\right) \vect{v}^{\rm sh}_{N},
\label{vnew}
\ee
where the first term on the right-hand side is required to be positive to prevent change of direction between $\vect{v}_{N+1}$ and $\vect{v}^{\rm sh}_{N}$. Orbit $N+1$ starts from the intersection point given by equation \eqref{rint}. Combined with its initial velocity, it allows to compute the orbital elements of orbit $N+1$.

\subsection{Equivalent differential equation}

\label{eqdif}

As the stream follows the succession of orbits described above, its energy $\epsilon$ and angular momentum $j$ vary. In the $\epsilon$-$j$ plane, the stream evolution is equivalent to the differential equation
\be
\frac{\diff j}{\diff \epsilon} = \frac{\Delta j}{\Delta \epsilon}
\label{djden}
\ee
as long as the number of ellipses describing the stream evolution is sufficiently large, where $\Delta \epsilon$ and $\Delta j$ are given by equations \eqref{deltaen} and \eqref{deltaj} respectively. Using the scaled quantities $\bar{\epsilon}=-\epsilon/c^2$ and $\bar{j}=j/(\rg c)$, equation \eqref{djden} becomes
\be
\frac{\diff \bar{j}}{\diff \bar{\epsilon}}=-2 \left(\frac{K_e}{e^2}\right) \left(\frac{v_{\rm A}}{v_{\rm c}}\right)^2 \frac{\bar{j}^3}{\sin^2(3 \pi /\bar{j}^2 )}.
\label{djdenbar}
\ee
In addition, the precession angle has been written as a function of angular momentum combining equations \eqref{angmom} and \eqref{precang}. This differential equation can be solved numerically for the initial conditions $\bar{\epsilon}_0$ and $\bar{j}_0$, obtained from equations \eqref{en0} and \eqref{j0}. The use of scaled quantities makes equation \eqref{djdenbar} independent of $\mh$ and $\beta$. However, the initial conditions depend on these parameters as $\bar{\epsilon}_0 \propto M^{1/3}_{\rm h}$ and $\bar{j}_0 \propto M^{-1/3}_{\rm h} \beta^{-1/2}$.

An analytical solution can be found by slightly modifying equation \eqref{djdenbar}. The small angle approximation $\sin \theta \approx \theta$ allows to simplify the denominator. In addition, since $K_e/e^2$ only varies by a factor of a few with $e$, it can be replaced by an average value $\tilde{K}\equiv\left<K_e/e^2\right>$. Numerically, we find that this factor can be fixed to $\tilde{K} = 5$ independently on the parameters. The resulting simplified equation is $\diff \bar{j}/\diff \bar{\epsilon} = -2 \tilde{K}  (\vaovc)^2 \bar{j}^7 / (9 \pi^2)$ whose analytical solution is

\be
\bar{\epsilon}-\bar{\epsilon}_0 = \frac{3\pi^2}{4\tilde{K}}  \left(\frac{v_{\rm A}}{v_{\rm c}}\right)^{-2} \left(\bar{j}^{-6}-\bar{j_0}^{-6}\right).
\label{eqsimp}
\ee
This simplified solution will be used in Section \ref{results} to prove interesting properties associated to the stream evolution.

\subsection{Stream evolution outcome}
\label{outcome}

As described above, the stream evolution is modelled by a succession of ellipses. The orbital elements of any orbit $N$ can be computed iteratively knowing the orbital changes between successive orbits. This iteration is stopped when the stream reaches one of the two following possible outcomes. They correspond to critical values of the orbit angular momentum and eccentricity, below which the computation is stopped.
\begin{enumerate}
\item \textit{Ballistic accretion}: if $j_N < \jacc \equiv 4\rg c$, the angular momentum of the stream is low enough for it to be accreted onto the black hole without circularizing.\\
\item \textit{Circularization}: if $e_N<e_{\rm circ}=1/3$, which corresponds to a stream apocentre equal to only twice its pericentre, we consider that the stream has circularized.
\end{enumerate}
Strictly speaking, the expression adopted for $\jacc$ is valid only for a test-particle on a parabolic orbit. For a circular orbit, it reduces to $2 \sqrt{3}\rg c$ \citep{hobson2006}, which is lower by a factor of order unity. However, this choice does not significantly affect our results as will be demonstrated in Section \ref{dynamical}. We therefore consider $\jacc$ as independent of the stream orbit. Our choice for the critical eccentricity $e_{\rm circ}$ can be understood by looking at the integral term in equation \eqref{ke}, below which the function $f_e$ is defined. Our stopping criterion $e<1/3$ implies $f_e(\pi)/f_e(0)<2$, which means that the stream loses less than twice as much angular momentum at apocentre than at pericentre. It is therefore legitimate to assume that angular momentum is lost homogeneously along the stream orbit from this point on.

If the computation ends with criterion (i), the stream is accreted. Its subsequent evolution is then irrelevant since it leads to no observable signal. If instead the computation ends with criterion (ii), a circular disc forms from the stream. This disc evolution is driven by magnetic stresses only, which act to shrink the disc nearly circular orbit until it reaches the innermost stable circular orbit, where it is accreted onto the black hole.

\section{Results}
\label{results}

We now present the results of our stream evolution model, which depends on three parameters: the black hole mass $\mh$, the penetration factor $\beta$ and the ratio of Alfvén to circular velocity $\vaovc$.\footnote{3D visualizations of the results presented in this paper can be found at \url{http://home.strw.leidenuniv.nl/~bonnerot/research.html}.} The first two parameters define the initial orbit of the debris through equations \eqref{en0} and \eqref{j0}, from which the iteration starts. As can be seen from equation \eqref{deltaj}, the parameter $\vaovc$ sets the efficiency of magnetic stresses at removing angular momentum from the stream. The star's mass and radius are fixed to the solar values.

The time required for the stream to reach a given orbital configuration is defined as the time spent by the most bound debris in all the previous orbits, starting from its first passage at pericentre after the disruption. Of particular importance is the time required for the stream to reach its final configuration, corresponding to either ballistic accretion or circularization. This evolution time is denoted $\tev$.\footnote{If the stream shrinks by a large factor from one orbit to the next, the debris might be distributed on several distinct orbits. It is then possible that the whole stream has not reached its final configuration even though the most bound debris did. This could lead to an underestimate of $\tev$.}

As in Section \ref{eqdif}, the scaled energy and angular momentum $\bar{\epsilon}=-\epsilon/c^2>0$ and $\bar{j}=j/(\rg c)$ will often be adopted in the following. Note that energy loss implies an increase of $\bar{\epsilon}$ due to the minus sign.

\subsection{Dynamical evolution of the stream}
\label{dynamical}

\begin{figure}
\epsfig{width=0.47\textwidth, file=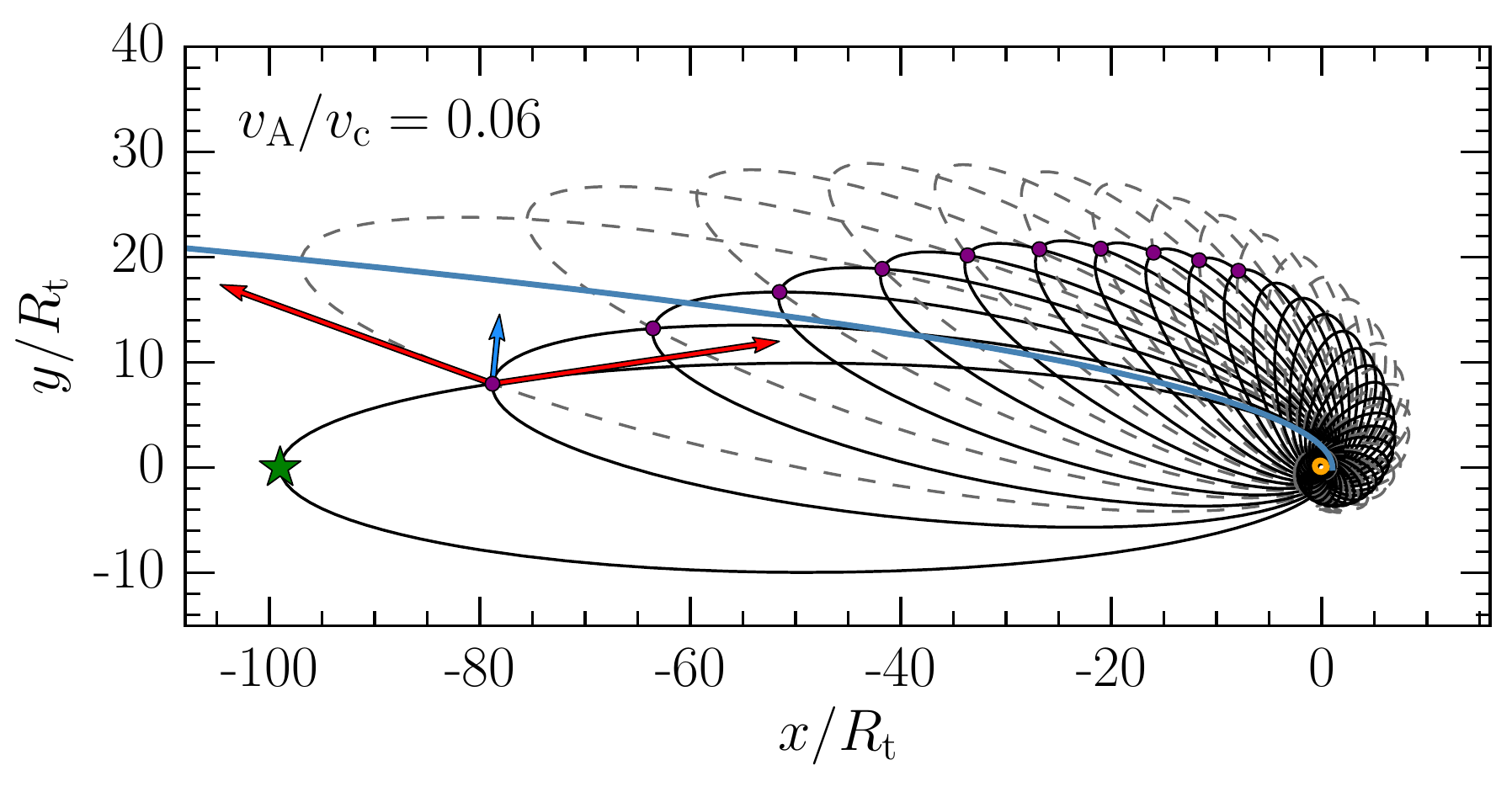}
\epsfig{width=0.47\textwidth, file=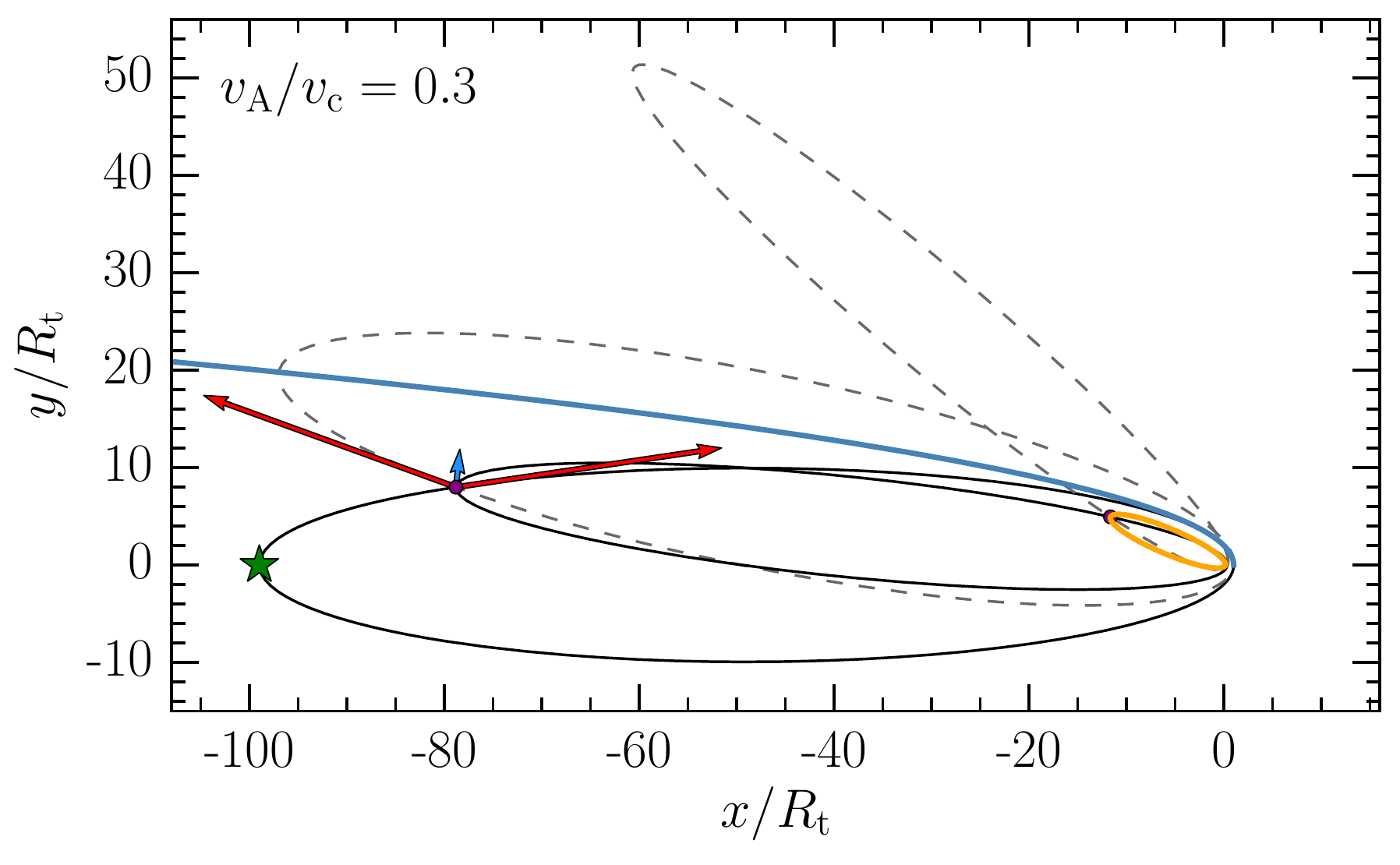}
\caption{Stream evolution for two magnetic stresses efficiencies $\vaovc= 0.06$ (upper panel) and 0.3 (lower panel). The black hole mass and penetration factor are fixed to $\mh=10^6 \msun$ and $\beta=1$. The black hole is at the origin. The succession of ellipses starts from the orbit of the most bound debris, whose apocentre is indicated by a green star on the left of the figure. Each stream orbit is divided into two ellipses. The stream elements moving towards the black hole follow the black ellipses. The dashed grey ellipses, precessed with respect to the black ones, are covered by the gas elements moving away from the black hole after pericentre passage. The intersection point is located where the black and grey ellipses cross. At this point, the orbit of the stream changes due to shocks and magnetic stresses. The stream elements then infall towards the black hole on the next solid black ellipse. The first ten self-crossing points are indicated by the purple dots. At the first crossing point, where the transition between orbit 0 and 1 happens, the red arrows show the velocity of the components involved in the associated shock, $\vect{v}^{\rm in}_0$ and $\vect{v}^{\rm out}_0$. The blue arrow indicates the initial velocity of orbit 1, $\vect{v}_1$, after the debris experienced both shocks and magnetic stresses. This situation is also illustrated in Fig. \ref{fig1} for $N=0$. The final orbit of the stream, for which one of the two stopping criteria is satisfied, is depicted in orange. For $\vaovc= 0.06$, the stream circularizes to form a disc. For $\vaovc= 0.3$, the stream is ballistically accreted before circularizing. The blue line represents a parabolic trajectory with pericentre $\rp$, equal to that of the star. The trajectories of the debris falling back towards the black hole within the tail are therefore contained between this line and the orbit of the most bound debris.}
\label{fig2}
\end{figure}

\begin{figure}
\epsfig{width=0.47\textwidth, file=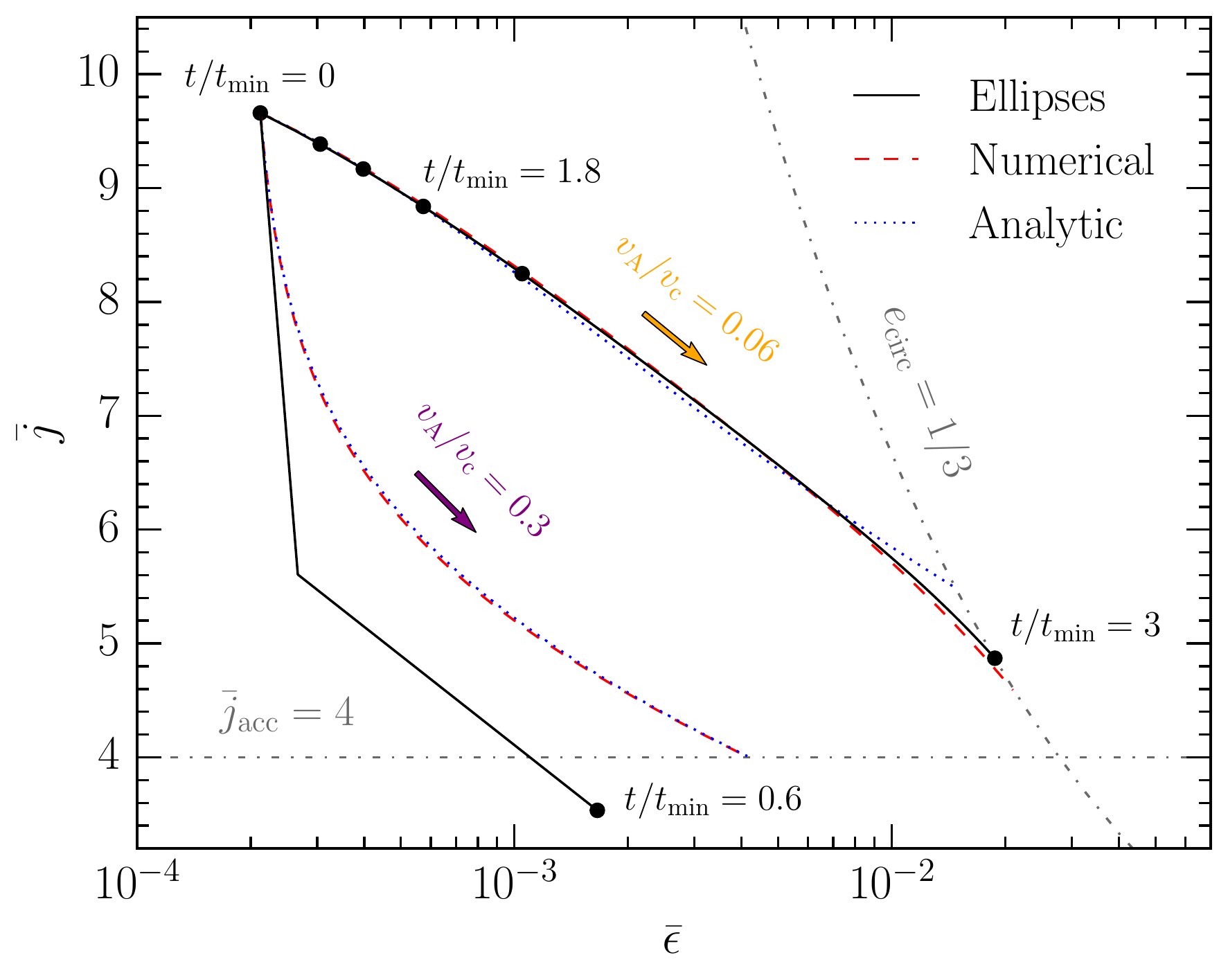}
\caption{Stream evolution shown in the $\bar{j}-\bar{\epsilon}$ plane for two values of $\vaovc = 0.06$ (orange arrow) and 0.3 (purple arrow). The black hole mass and penetration factor are fixed to $\mh=10^6 \msun$ and $\beta=1$. The spatial stream evolution for these two examples is shown in Fig. \ref{fig2}. The black solid line corresponds to the succession of ellipses. The red dashed line shows the numerical solution of the differential equation \eqref{djden} while the blue dotted line shows the simplified analytical solution given by equation \eqref{eqsimp} with $\tilde{K} = 5$. The horizontal grey dash-dotted line represents the angular momentum $\bar{j}_{\rm acc}=4$ below which ballistic accretion occurs while the vertical one shows the eccentricity $e_{\rm crit} = 1/3$ below which circularization happens.}
\label{fig3}
\end{figure}

We start by investigating the stream evolution for a tidal disruption by a black hole of mass $\mh = 10^6 \msun$ with a penetration factor $\beta=1$. Two different magnetic stresses efficiencies are examined, corresponding to $v_{\rm A}/v_{\rm c}= 0.06$ and 0.3. The stream evolution is shown in Fig. \ref{fig2} for these two examples. It is represented by the ellipses it goes through, starting from the orbit of the most bound debris whose apocentre is indicated by a green star. The final configuration of the stream is shown in orange. For $v_{\rm A}/v_{\rm c}= 0.06$ (upper panel), the stream gradually shrinks and becomes circular at $\tev/\tmin = 3$. The evolution differs for $v_{\rm A}/v_{\rm c}= 0.3$ (lower panel) where the stream ends up being ballistically accreted at $\tev/\tmin=0.6$. These evolutions can also be examined using Fig. \ref{fig3} (black solid lines), which shows the associated path in the $\bar{j}-\bar{\epsilon}$ plane. For $v_{\rm A}/v_{\rm c}= 0.06$ (orange arrow), the stream evolves slowly initially as can be seen from the black points associated to fixed time intervals. As it loses more energy and angular momentum, the evolution accelerates and the stream rapidly circularizes reaching the grey dash-dotted line on the right of the figure that corresponds to $e=e_{\rm circ}$. Note that if no magnetic stresses were present, the stream would still circularize but following an horizontal line in this plane. For $v_{\rm A}/v_{\rm c}= 0.3$ (purple arrow), the stream rapidly loses angular momentum which leads to its ballistic accretion when $j<\jacc$, crossing the horizontal grey dash-dotted line. The stream evolution outcome therefore depends on the efficiency of magnetic stresses, given by the parameter $\vaovc$. If they act fast enough, the stream loses enough angular momentum to be accreted with a substantial eccentricity. Otherwise, the energy loss through shocks dominates, resulting in the stream circularization. \\

\begin{figure}
\epsfig{width=0.47\textwidth, file=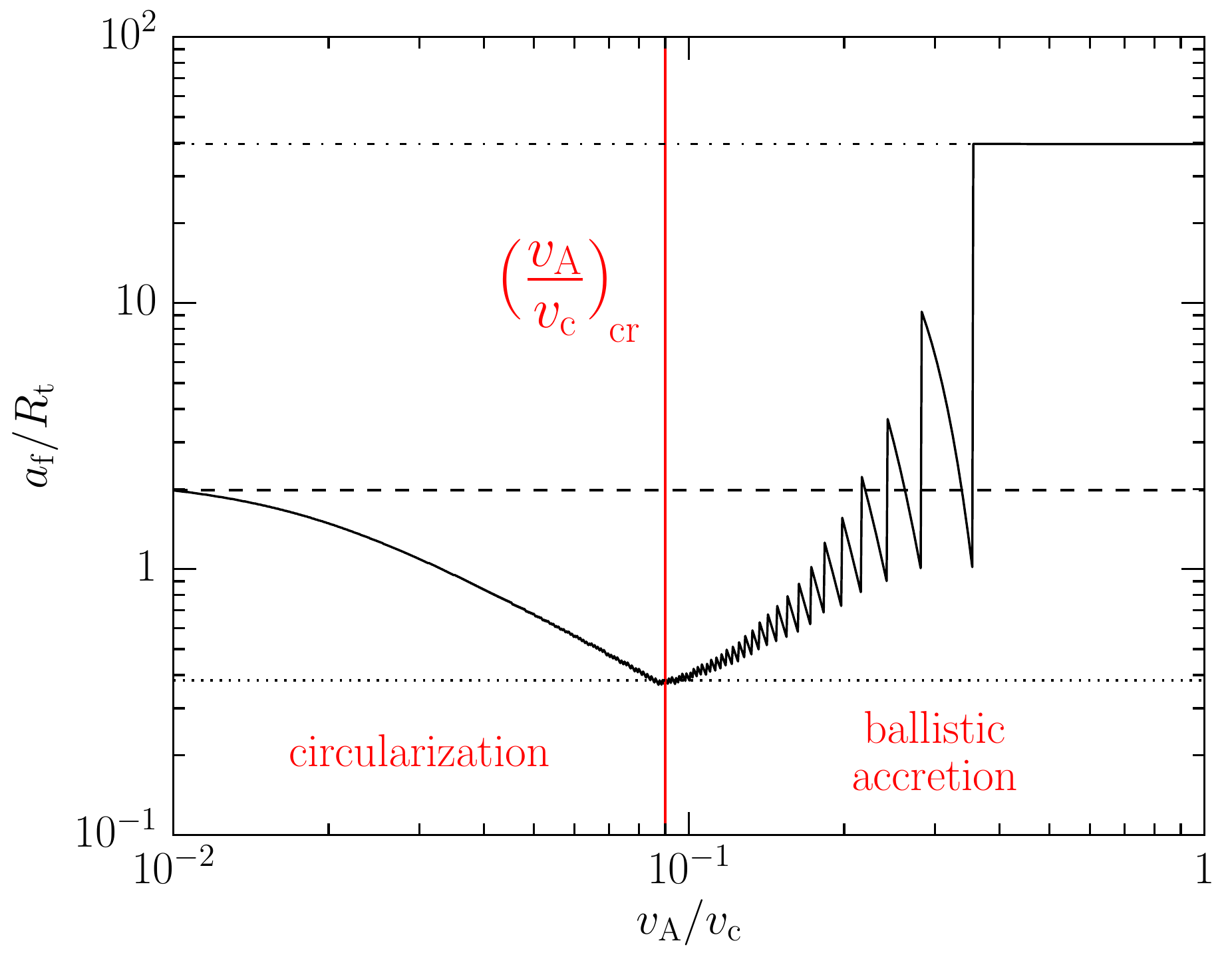}
\caption{Semi-major axis of the stream at the end of the stream evolution as a function of the magnetic stresses efficiency $\vaovc$. The two other parameters are fixed to $\mh=10^6 \msun$ and $\beta=1$, for which the spatial stream evolution is shown in Fig. \ref{fig2}. From top to bottom, the horizontal lines have the following meanings. The dot-dashed line indicates the semi-major axis of the stream after the first shock $a_1 \approx R^{\rm int}_0/2 = 40 \rt$ where $R^{\rm int}_0$ denotes the distance to the first self-crossing point. The dashed line represents the circularization radius obtained from angular momentum conservation $(1+e_0)R^{\rm p}_0 \approx 2 \rt$. Finally, the dotted line shows the semi-major axis corresponding to a circular orbit with angular momentum $\jacc$, equal to $18 \rg \approx 0.4 \rt$.}
\label{fig4}
\end{figure}

\begin{figure}
\epsfig{width=0.47\textwidth, file=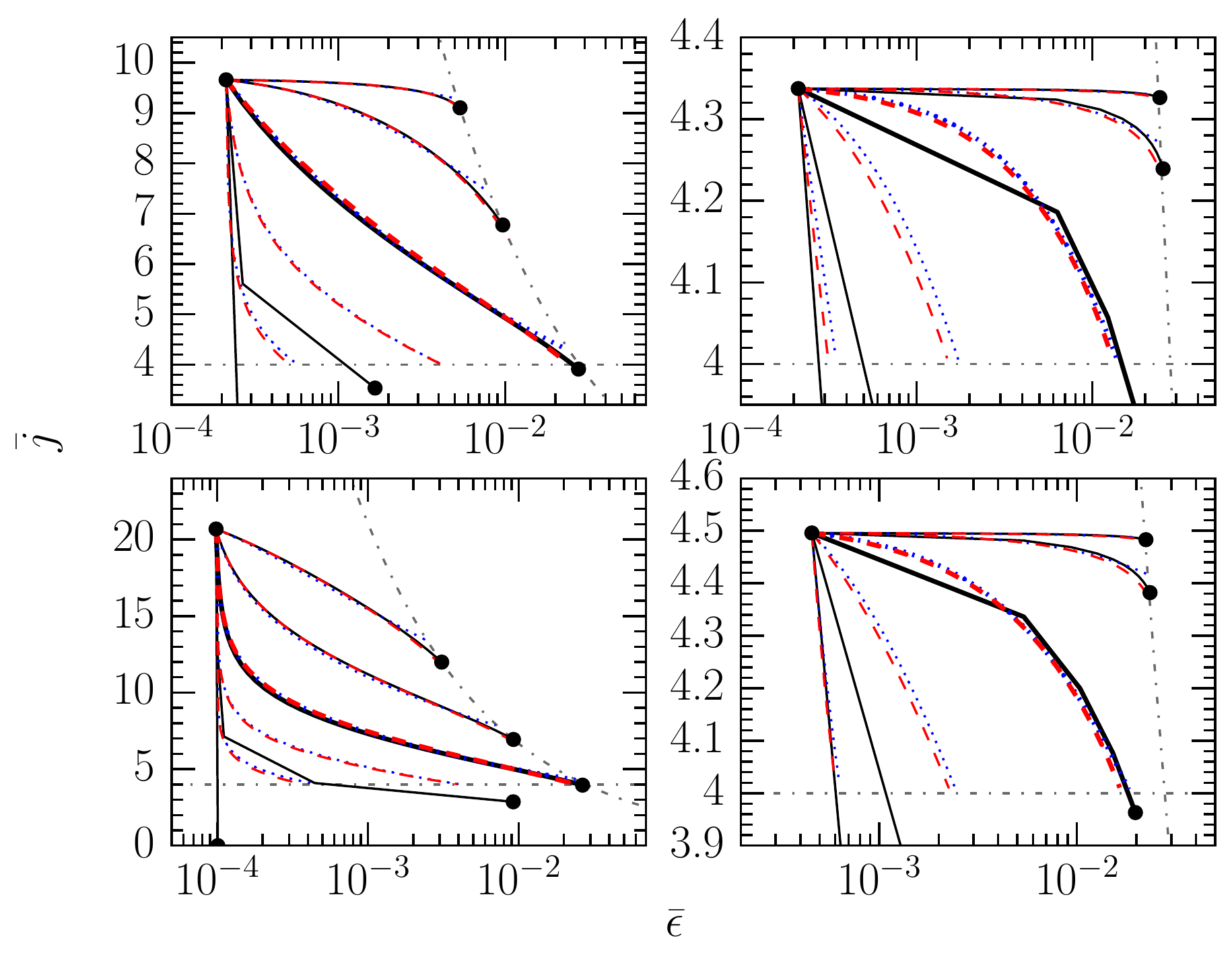}
\caption{Stream evolution shown in the $\bar{j}-\bar{\epsilon}$ plane for various values of the parameters. The meaning of the different lines are the same as in Fig. \ref{fig3}. In each panel, the five sets of curves correspond to different values of the magnetic efficiency $\vaovc = 0.01$, 0.03, 0.1, 0.3 and 1 (from top to bottom). The different panels correspond to different values of $\mh$ and $\beta$. The two top panels show $\mh=10^6 \msun$ for $\beta=1$ (upper left) and for $\beta=5$ (upper right). The two bottom panels adopt $\beta=1$ for $\mh=10^5 \msun$ (lower left) and $\mh=10^7 \msun$ (lower right). The set of thicker curves is associated to a magnetic stresses efficiency $\vaovc=0.1$, which approximately corresponds to the critical value (equation \eqref{vaovccr}) defining the boundary between circularization and ballistic accretion, independently on $\mh$ and $\beta$.}
\label{fig5}
\end{figure}

The influence of the magnetic stresses efficiency $\vaovc$ on the stream evolution can be analysed more precisely from Fig. \ref{fig4}, which shows the semi-major axis $a_{\rm f}$ of the stream at the end of its evolution as a function of $\vaovc$. The other parameters are fixed to $\mh=10^6 \msun$ and $\beta=1$. For low values of $\vaovc \approx 10^{-2}$, the stream circularizes at the circularization radius $(1+e_0)R^{\rm p}_0 \approx 2 \rt$ obtained from angular momentum conservation (horizontal dashed line). As $\vaovc$ increases, the stream circularizes closer to the black hole since its angular momentum decreases due to magnetic stresses during the circularization process. At $\vaovc \approx 10^{-1}$, the final semi-major axis reaches its lowest value. This minimum corresponds to circularization with an angular momentum exactly equal to $\jacc$, for which $a_{\rm f} = 18 \rg \approx 0.4 \rt$. For $\vaovc \gtrsim 10^{-1}$, the stream ends its evolution by being ballistically accreted. This demonstrates again the existence of a critical value $(\vaovc)_{\rm cr}$ for the magnetic stresses efficiency (vertical solid red line) that defines the boundary between circularization (on the left) and ballistic accretion (on the right). The final semi-major axis reaches a plateau at $\vaovc \gtrsim 0.4$, for which the stream gets ballistically accreted after its first shock. In this region, $a_{\rm f} \approx R^{\rm int}_0/2 = 40 \rt$, where $R^{\rm int}_0 = 80 \rt$ denotes the distance to the first intersection point (see Fig. \ref{fig2}). The oscillations visible for $\vaovc \lesssim 0.4$ are associated to different numbers of orbits followed by the stream before its ballistic accretion. On the left end of the plateau, the stream gets accreted after the first self-crossing shock with an angular momentum just below $\jacc$. Decreasing $\vaovc$ by a small amount prevents this ballistic accretion since the stream now has an angular momentum just above $\jacc$ after the first shock. The stream therefore undergoes a second shock, which is strong since the previous pericentre passage occurred close to the black hole with a large precession angle. The stream semi-major axis therefore decreases by a large amount before ballistic accretion. This results in a discontinuity in $a_{\rm f}$ at the edge of the plateau. Decreasing $\vaovc$ further, the stream passes further away from the black hole which reduces apsidal precession and weakens the second shock. As a result, $a_{\rm f}$ increases. When $\vaovc$ becomes low enough for ballistic accretion to be prevented after the second shock, a strong third shock occurs before ballistic accretion which causes a second discontinuity due to the sharp decrease of $a_{\rm f}$. The same mechanism occurs for larger numbers of orbits preceding ballistic accretion, producing the other discontinuities and increases of $a_{\rm f}$ seen for a decreasing $\vaovc$ and resulting in this oscillating pattern.

The role of $\vaovc$ in determining the stream evolution outcome can be understood by looking again at Fig. \ref{fig3}. The black solid lines are associated to the succession of ellipses described above. The red dashed line shows the numerical solution of the equivalent differential equation \eqref{djdenbar} while the blue dotted line corresponds to the simplified analytical version, given by equation \eqref{eqsimp}. For $\vaovc=0.06$, these three descriptions are consistent and able to capture the stream evolution. For $\vaovc=0.3$, the evolution obtained from the succession of ellipses differs from the two others. This is expected since the stream goes through only three ellipses in this case, not enough for its evolution to be described by the equivalent differential equation. An interesting property of these solutions can be identified from equation \eqref{eqsimp}. As soon as $\bar{\epsilon} \gg \bar{\epsilon}_0$ and $\bar{j}^6 \ll \bar{j}^6_0$, the position of the stream in the $\bar{j}-\bar{\epsilon}$ plane becomes independent of the initial conditions $\bar{\epsilon}_0$ and $\bar{j}_0$, given by equations \eqref{en0} and \eqref{j0} respectively. It is therefore dependent on $\vaovc$ only, but not on $\mh$ and $\beta$ anymore. In this case, one expect the critical value $(\vaovc)_{\rm cr}$ of the magnetic stresses efficiency to also be completely independent of $\mh$ and $\beta$. In practice, the first condition $\bar{\epsilon} \gg \bar{\epsilon}_0$ is always satisfied as long as the stream loses energy by undergoing a few shocks since the initial orbit is nearly parabolic with $\bar{\epsilon}_0 \approx 0$. The second one $\bar{j}^6 \ll \bar{j}^6_0$ is however not satisfied in general for low values of $\bar{j}_0$. In fact, $\bar{j}_0$ can be already close to $\bar{j}_{\rm acc}=4$ for large $\beta$ or $\mh$, since $\bar{j}_0 = j_0/(\rg c) \propto \beta^{-1/2} \mh^{-1/3}$ (equation \eqref{j0}). To account for this possibility, we define the factor $f_0 \equiv (\bar{j}_0/\bar{j}_{\rm acc})^{-6}$ satisfying $0<f_0<1$. The critical value $(\vaovc)_{\rm cr}$, for which the stream circularizes with an angular momentum $\bar{j} = 4$ (see Fig. \ref{fig4}), can then be obtained analytically by fixing $e=1/3$ and $\bar{j}=4$ in equation \eqref{eqsimp} combined with $1-e^2=2\bar{j}^2\bar{\epsilon}$.  This yields $(\vaovc)_{\rm cr} = \pi (4096\tilde{K}/27(1-f_0))^{-1/2}$ whose numerical value is
\be
\left(\frac{v_{\rm A}}{v_{\rm c}}\right)_{\rm cr} \approx 10^{-1} \left(1-f_0\right)^{1/2}
\label{vaovccr}
\ee
using $\tilde{K} = 5$. As anticipated, $(\vaovc)_{\rm cr} \approx 10^{-1}$ independently of $\mh$ and $\beta$ as long as $f_0 \ll 1$. This is the case for $\mh=10^6$ and $\beta=1$, for which $f_0 \approx 5 \times 10^{-3} \ll1$. We therefore recover the value of $(\vaovc)_{\rm cr} \approx 10^{-1}$ indicated in Fig. \ref{fig4} (red vertical line). For larger $\mh$ or $\beta$, the condition $f_0 \ll 1$ is not necessarily satisfied. For example, $f_0 \approx 0.5$ for $\mh=10^7$ and $\beta=1$. In this case, $(\vaovc)_{\rm cr}$ is slightly lower according to equation \eqref{vaovccr}, but only by a factor less than 2. In practice, $f_0 \approx 1$ only in the extreme case where the stream is originally on the verge of ballistic accretion with $j_0$ very close to $\jacc$. We can therefore conclude that the magnetic stresses efficiency, delimiting the boundary between circularization and ballistic accretion, has a value $(\vaovc)_{\rm cr} \approx 10^{-1}$ largely independently on the other parameters of the model, $\mh$ and $\beta$. In addition, note that this value is not significantly affected by the choice we made for $\jacc$ as mentioned in Section \ref{outcome}.

The value of $(\vaovc)_{\rm cr}$ derived analytically in equation \eqref{vaovccr} can be confirmed from Fig. \ref{fig5}, which shows the stream evolution in the $\bar{j}-\bar{\epsilon}$ plane for various values of the parameters. The different lines have the same meaning as in Fig. \ref{fig3}. In each panels, the six sets of lines shows different values of $\vaovc = 0.01$, 0.03, 0.1, 0.3 and 1 (from top to bottom). The different panels correspond to various choices for $\mh$ and $\beta$. The thick set of lines indicates $\vaovc = 10^{-1}$. As expected from equation \eqref{vaovccr}, it corresponds exactly to the boundary between circularization and ballistic accretion for $\mh =10^6 \msun$ (upper left panel) and $10^5 \msun$ (lower left panel), both with $\beta =1$. This is because $f_0 \ll 1$ in these cases. However, increasing the black hole mass to $\mh =10^7 \msun$ (lower right panel) or the penetration factor to $\beta=5$ (upper right panel), the stream is ballistically accreted for $\vaovc = 10^{-1}$. This comes from the fact that $f_0$ is not completely negligible in these cases, which implies $(\vaovc)_{\rm cr} < 10^{-1}$ according to equation \eqref{vaovccr}.

\begin{figure}
\epsfig{width=0.47\textwidth, file=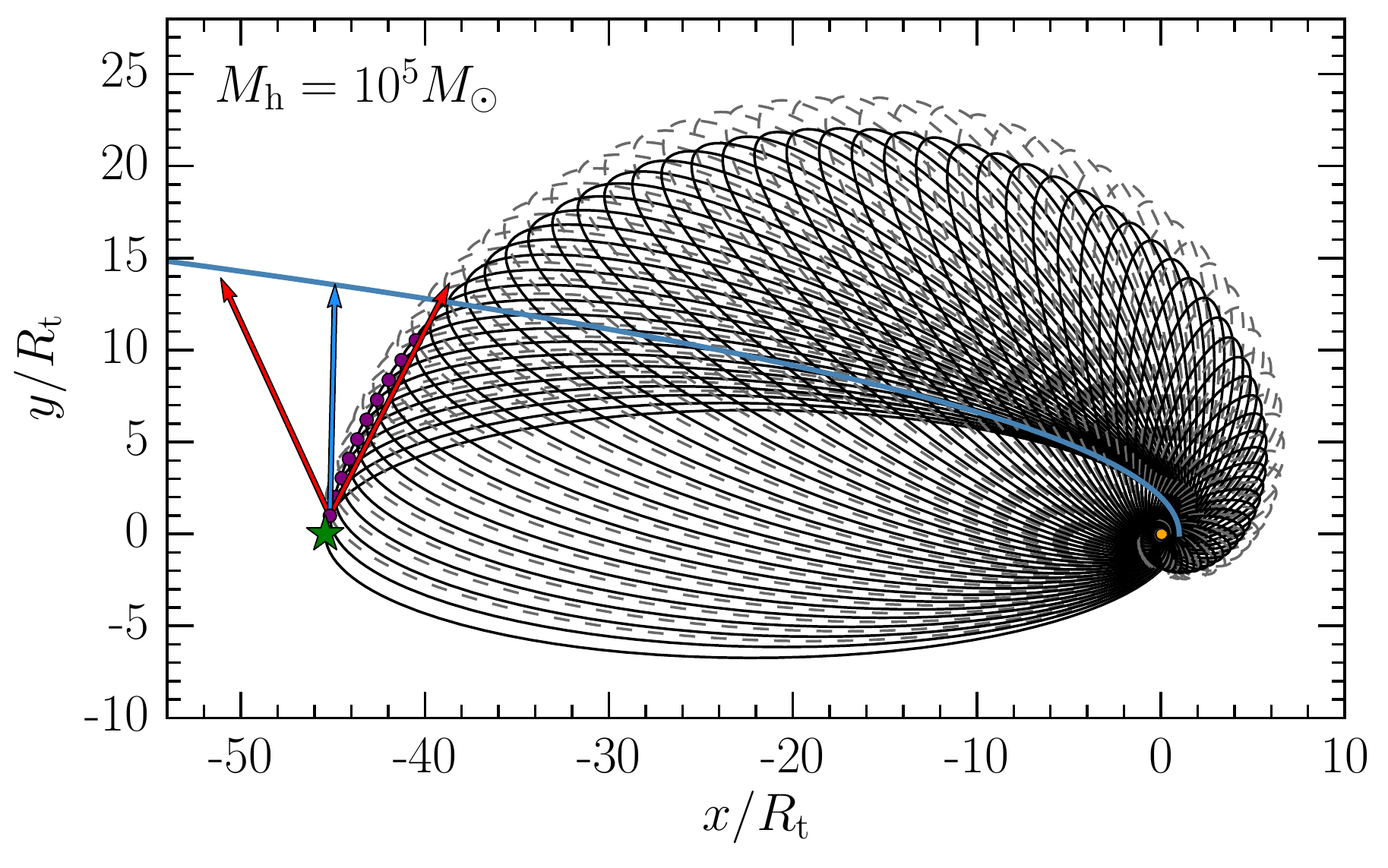}
\epsfig{width=0.47\textwidth, file=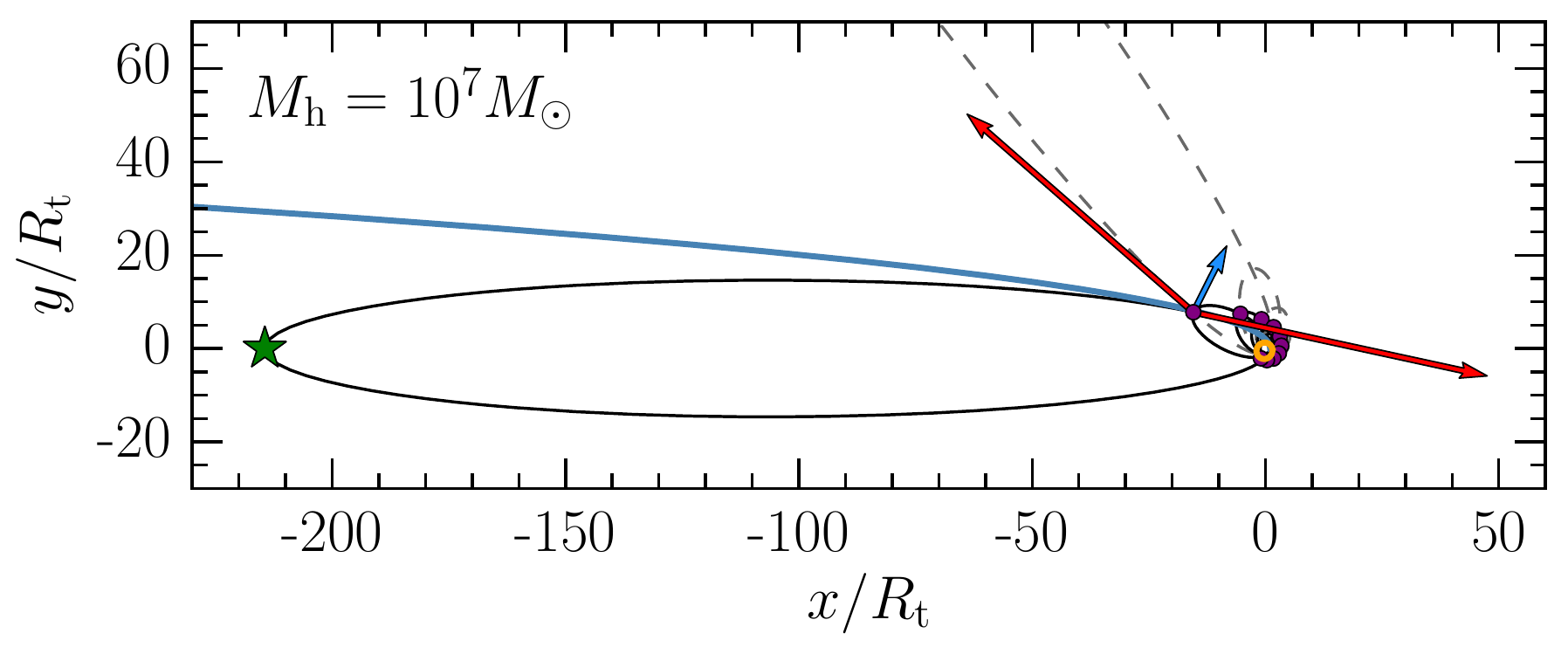}
\caption{Stream evolution for $\mh=10^5\msun$ (upper panel) and $\mh=10^7\msun$ (lower panel) with $\beta=1$ and $\vaovc=0.06$. The different elements of this figure have the same meaning as in Fig \ref{fig2}, whose upper panel shows the intermediate case with $\mh=10^6\msun$.}
\label{fig6}
\end{figure}

\begin{figure}
\epsfig{width=0.47\textwidth, file=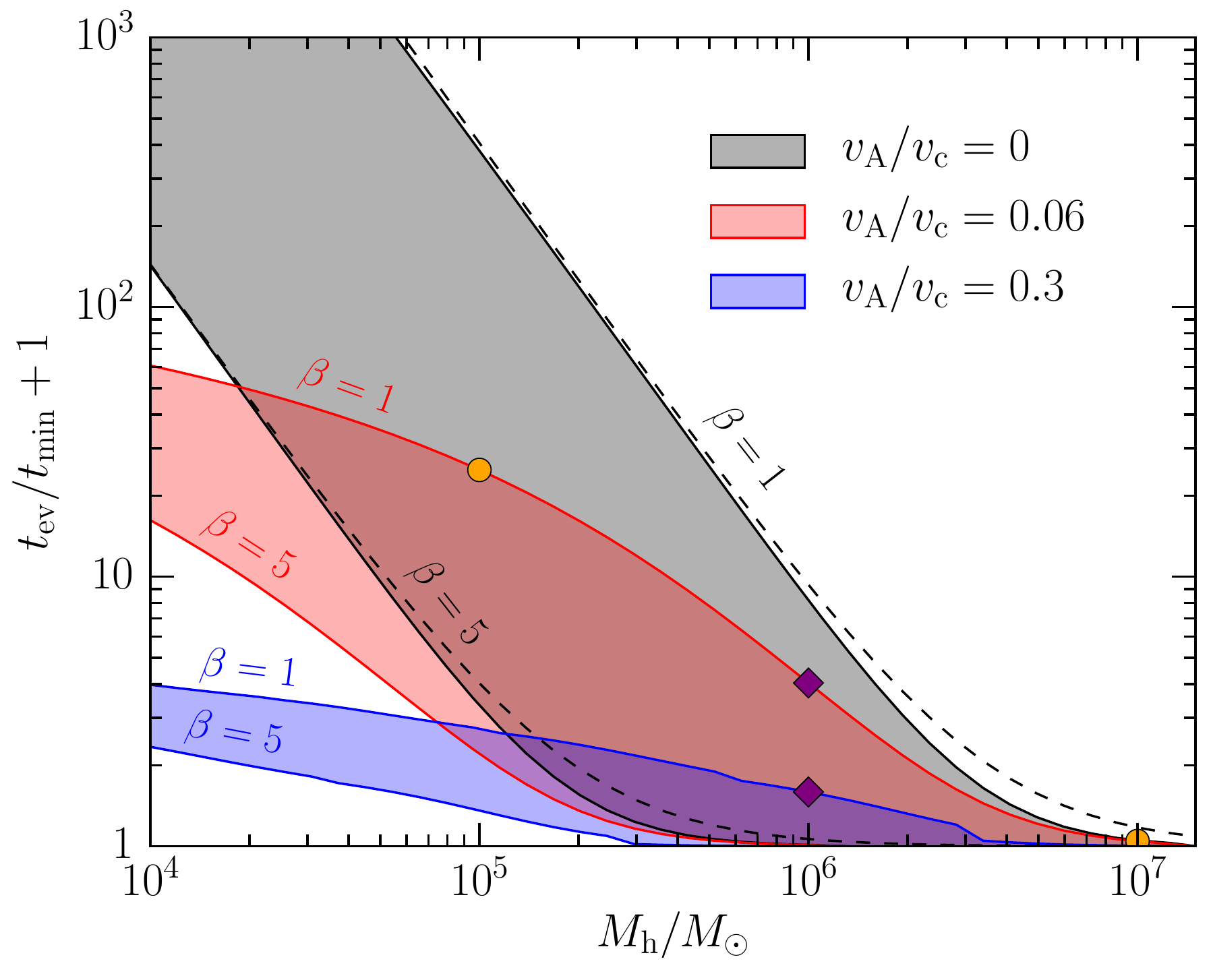}
\caption{Evolution time of the stream as a function of black hole mass. The three shaded areas correspond to three values of the magnetic stresses efficiency $\vaovc=0$ (black), 0.06 (red) and 0.3 (blue). Each area is delimited by two lines, which are associated to $\beta=1$ (upper line) and 5 (lower line). The dashed black lines show the analytical estimate for the evolution time in the absence of magnetic stresses given by equation \eqref{tevana} for $\beta=1$ (top line) and 5 (bottom line). The cases shown in Fig. \ref{fig2} and \ref{fig6} are represented by the two purple diamonds and orange circles respectively.}
\label{fig7}
\end{figure}

Although the stream evolution outcome only varies with the magnetic efficiency $\vaovc$, the time required to reach this final configuration and the orbits it goes through in the process are dependent on $\mh$ and $\beta$ in addition to $\vaovc$. The effect of varying the black hole mass only can be seen by looking at Fig. \ref{fig6}, which shows the stream evolution for $\mh=10^5 \msun$ (upper panel) and $10^7 \msun$ (lower panel) keeping the other two parameters fixed to $\beta=1$ and $\vaovc=0.06$. The intermediate case, with $\mh=10^6 \msun$, is shown in Fig. \ref{fig2} (upper panel). For larger black hole masses, the time for the stream to circularize is shorter, varying from $\tev/\tmin=24$ to 0.05 from $\mh=10^5\msun$ to $10^7\msun$. Note that this time is also reduced in physical units, from $\tev = 310$ to 6 days. The reason is that increasing the black hole mass leads to a larger precession angle, which causes the stream to self-cross closer to the black hole and lose more energy. As a result, the stream evolves faster to its final configuration. The same trend is expected if the penetration $\beta$ is increased since the precession angle scales as $\phi \propto \rg/\rp \propto \beta \mh^{2/3}$. Fig. \ref{fig7} proves this fact by showing the evolution time as a function of black hole mass for several values of $\beta$ and $\vaovc$. The different colours correspond to different values of $\vaovc$ while the width of the shaded areas represents various values of $\beta$ from 1 (upper line) to 5 (lower line). Furthermore, it can be seen that the stream evolves more rapidly for larger values of $\vaovc$. For example, $\tev$ decreases by about 2 orders of magnitude for $\mh=10^5 \msun$ when the magnetic efficiency is increased from $\vaovc=0$ to 0.3. This is because angular momentum loss from the stream at apocentre causes a decrease of its pericentre distance, which results in stronger shocks and a faster stream evolution.\\

For $\vaovc=0$, the angular momentum of the stream is conserved. The energy lost by the stream at each self-crossing is then independent of the stream orbit as explained at the end of Section \ref{shocks}. In this case, the evolution time obeys the simple analytic expression
\be
\frac{\tev}{\tmin} = \frac{2  \Delta \epsilon}{ \Delta \epsilon_0} = 8.3 \left( \frac{\mh}{10^6 \msun}\right)^{-5/3} \beta^{-3},
\label{tevana}
\ee
where $\Delta \epsilon_0$ represents the energy lost at each self-crossing shock, given by equation \eqref{deltaenzero}. It should not be confused with $\Delta \epsilon$, which is the initial energy of the stream equal to that of the most bound debris according to equation \eqref{en0}. For clarity, the derivation of equation \eqref{tevana} is made in Appendix A. As can be seen from Fig. \ref{fig7}, this analytical estimate (dashed black lines) matches very well the value of the evolution time obtained from the succession of ellipses with $\vaovc = 0$ (black solid lines) for both $\beta=1$ and 5. Equation \eqref{tevana} comes from a mathematical derivation but does not have a clear physical reason. Imposing $\tev/\tmin$ to be constant leads to the relation $\beta \propto M^{-5/9}_{\rm h}$. Interestingly, this dependence is similar although slightly shallower than that obtained by imposing $\rp/\rg$ to be constant, which gives $\beta \propto M^{-2/3}_{\rm h}$. This latter relation has been used by several authors to extrapolate the results of disc formation simulations from unphysically low-mass black holes to realistic ones \citetext{e.g. \citealt{shiokawa2015}}.

When $\vaovc>(\vaovc)_{\rm cr} \approx 10^{-1}$ and the stream is eventually ballistically accreted, the evolution time $\tev$ is always less than a few $\tmin$ for $\mh \approx 10^6\msun$. Moreover, a significant amount of energy is lost before accretion, resulting in a final orbit substantially less eccentric than initially. A typical case of ballistic accretion is illustrated by Fig. \ref{fig2} (lower panel). The only scenario where significant energy loss is avoided is if the stream is accreted immediately after the first shock. However, in this case, $\tev$ is very low. This behaviour is quite different from the evolution described by \citet{svirski2015}, for which the stream remains highly elliptical for tens of orbits progressively losing angular momentum via magnetic stresses before being ballistically accreted. Our calculations demonstrate instead that ballistic accretion happens on a short timescale, most of the time associated with a significant energy loss via shocks.\\

\begin{figure*}
\epsfig{width=0.47\textwidth, file=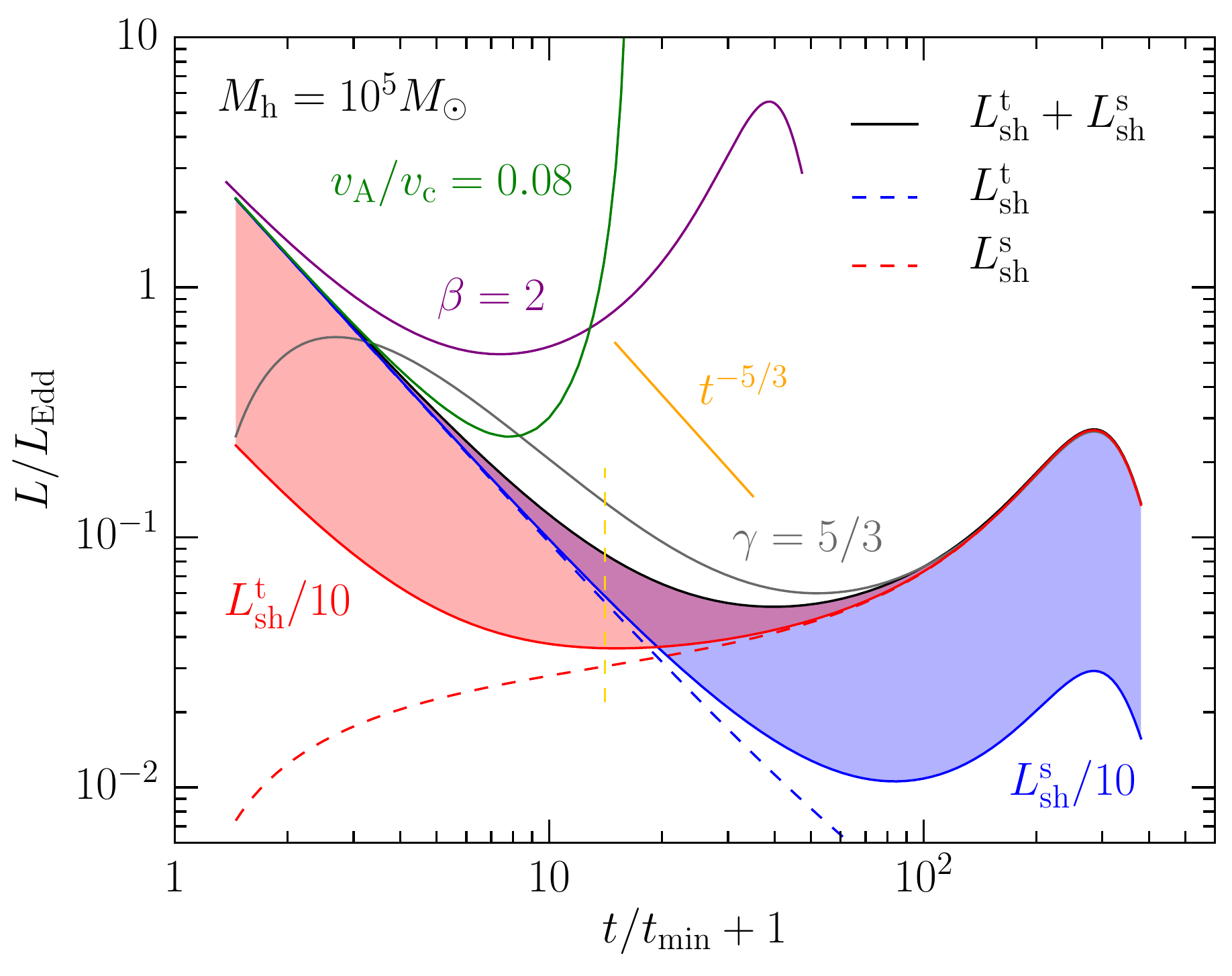}
\epsfig{width=0.47\textwidth, file=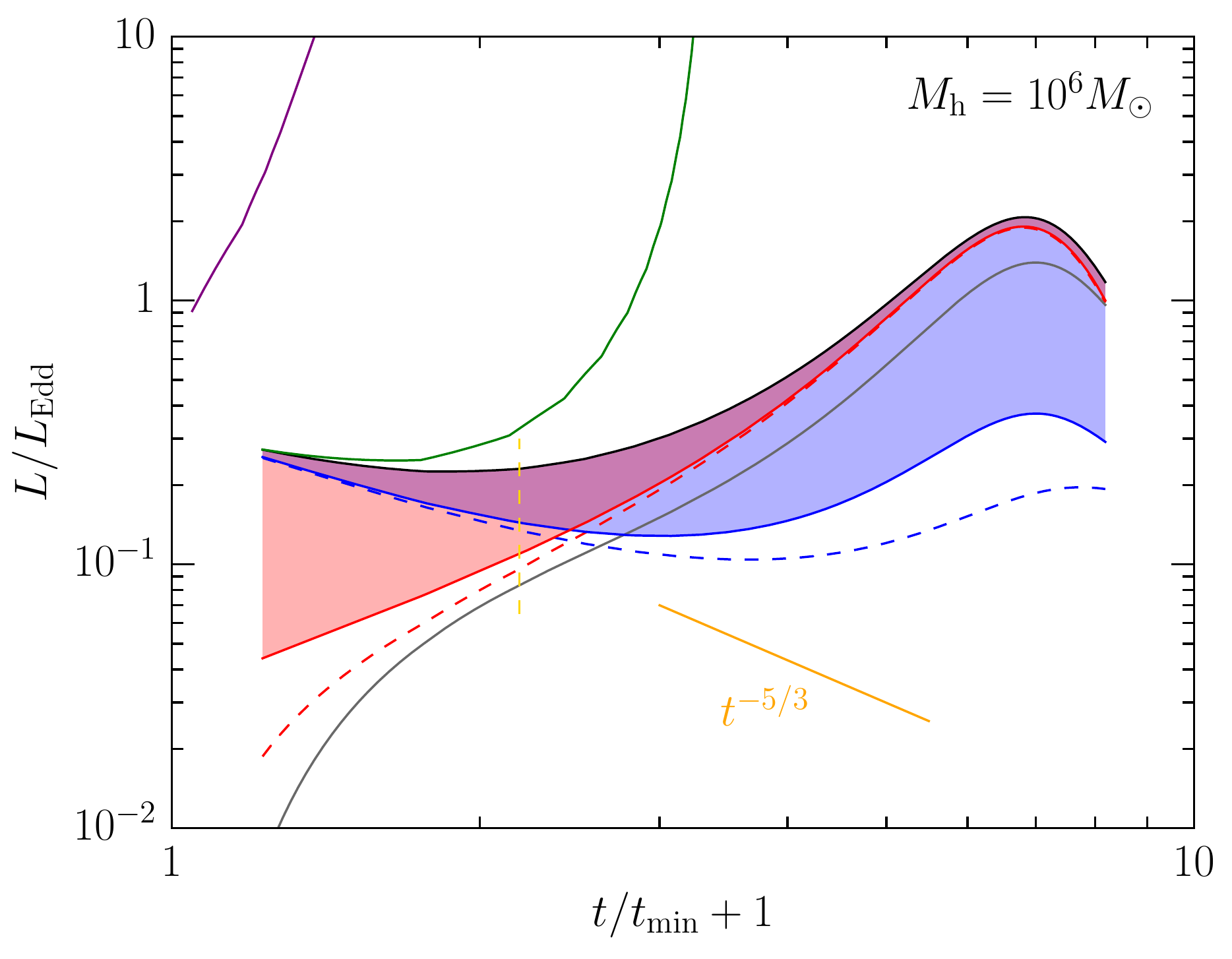}
\caption{Evolution of the stream shock luminosity $L^{\rm s}_{\rm sh}$ (red dashed line), tail shock luminosity $L^{\rm t}_{\rm sh}$ (blue dashed line), given by equations \eqref{lshs} and \eqref{lsht} respectively, and total shock luminosity $L^{\rm s}_{\rm sh}+L^{\rm t}_{\rm sh}$ (solid black line) for black hole masses $\mh=10^5\msun$ (left panel) and $10^6\msun$ (right panel) assuming a flat energy distribution for the fallback rate and a stream dissipation timescale ${\Delta t}_{\rm dis} = P_{\rm s}$ in equation \eqref{mdots}. Equal radiative efficiencies are adopted for the two shock sources, with $\eta^{\rm s}_{\rm sh} = \eta^{\rm t}_{\rm sh}= 1$. The other parameters are fixed to $\beta=1$ and $\vaovc=0$. The total luminosity is also shown for $\beta=2$ (purple solid line), $\vaovc=0.08$ (green solid line) and using the more precise fallback rate evolution of \citet{lodato2009} for a polytropic star with $\gamma=5/3$ (grey solid line) keeping the other parameters fixed. The shaded regions show the areas covered by the total luminosity as $L^{\rm t}_{\rm sh}$ (red area) and $L^{\rm s}_{\rm sh}$ (blue area) are decreased up to a factor of 10 (red and blue solid lines). All the luminosities are scaled by the Eddington value $\ledd$. The orange solid segment indicates the $t^{-5/3}$ slope. The vertical yellow dashed segment marks the time before which the locations of self-crossing and tail shocks remain similar, implying comparable radiative efficiencies $\eta^{\rm s}_{\rm sh} \approx \eta^{\rm t}_{\rm sh}$.}
\label{fig8}
\end{figure*}

\subsection{Observational appearance}
\label{observational}

We now investigate the main observational features associated to the stream evolution. Two sources of luminosity are identified, which can be evaluated from the dynamical stream evolution presented in Section \ref{dynamical}. The first source is associated to the energy lost by the stream due to self-intersecting shocks. The associated stream self-crossing shock luminosity can be evaluated as
\be
L^{\rm s}_{\rm sh}= \eta^{\rm s}_{\rm sh} \dot{M}_{\rm s} \Delta \epsilon_{\rm s},
\label{lshs}
\ee
where $\Delta \epsilon_{\rm s}$ is the instantaneous energy lost from the stream, obtained from the succession of orbits described in Section \ref{dynamical} via a linear interpolation between successive orbits. $\dot{M}_{\rm s}$ represents the mass rate at which the stream enters the shock, obtained from
\be
\dot{M}_{\rm s}=M_{\rm s}/{\Delta t}_{\rm dis},
\label{mdots}
\ee
where $M_{\rm s}$ is the mass of debris present in the stream and ${\Delta t}_{\rm dis}$ denotes the time required for all this matter to go through the shock and dissipate its orbital energy. We explore two different ways of computing the mass of the stream. The first assumes a flat energy distribution within the disrupted star leading to $M_{\rm s} = 0.5 M_{\star} (1- (t/\tmin + 1)^{-2/3})$. The second follows \citet{lodato2009}, which adopts a more precise description of the internal structure of the star, modelled by a polytrope. This latter approach results in a shallower increase of the stream mass. If all the gas present in the stream is able to pass through the intersection point, the time ${\Delta t}_{\rm dis}$ during which the debris energy is dissipated is equal to the orbital period of the stream $P_{\rm s}$. However, this can be prevented if a shock component, either the infalling or the ouflowing part of the stream, gets exhausted earlier than the other. In this case, part of the stream material keeps its original energy. Nevertheless, this gas will eventually join the rest of the stream and release its energy, only at slightly later times. This effect can therefore be accounted for by setting ${\Delta t}_{\rm dis} > P_{\rm s}$ by a factor of a few. Finally, the parameter $\eta^{\rm s}_{\rm sh}$ is the shock radiative efficiency, which accounts for the possibility that not all the thermal energy injected in the stream via shocks can be radiated away and participate to the luminosity $L^{\rm s}_{\rm sh}$. Its value depends on the optical thickness of the stream at the shock location and can be estimated by
\be
\eta^{\rm s}_{\rm sh}=\mathrm{min} (1,t^{\rm s}_{\rm sh}/t^{\rm s}_{\rm dif}),
\label{etashs}
\ee
$t^{\rm s}_{\rm dif}$ being the diffusion time at the self-crossing shock location while $t^{\rm s}_{\rm sh}$ denotes the duration of the shock, equal to the dynamical time at this position.

The second luminosity component is associated to the tail of gas constantly falling back towards the black hole. This newly arriving material inevitably joins the stream from an initially nearly radial orbit. During this process, its orbital energy decreases from almost zero to the orbital energy of the stream. The tail orbital energy lost is transferred into thermal energy via shocks and can be radiated. The associated tail shock luminosity is given by
\be
L^{\rm t}_{\rm sh}= \eta^{\rm t}_{\rm sh} \dot{M}_{\rm fb} \epsilon_{\rm s},
\label{lsht}
\ee
where $\epsilon_{\rm s}$ is the instantaneous energy of the stream obtained from the succession of ellipses by linearising between orbits. $\dot{M}_{\rm fb}$ is the mass fallback rate at which the tail reaches the stream. As for $\dot{M}_{\rm s}$ in equation \eqref{lshs}, we investigate two methods to compute the fallback rate. Assuming a flat energy distribution within the disrupted star gives $\dot{M}_{\rm fb}=1/3(M_{\star}/\tmin)(t/\tmin+1)^{-5/3}$, which corresponds to a fallback rate peaking when the first debris reaches the black hole, at $t=0$, and immediately decreasing as $t^{-5/3}$. Taking into account the stellar structure following \citet{lodato2009} leads to an initial rise of the fallback rate towards a peak, reached for $t$ of a few $\tmin$, followed by a decrease as $t^{-5/3}$ at later times. The parameter $\eta^{\rm t}_{\rm sh}$ is the radiative efficiency at the location of the shock between tail and stream present for the same reason as in equation \eqref{lshs}. It can be evaluated as in equation \eqref{etashs} by
\be
\eta^{\rm t}_{\rm sh}=\mathrm{min} (1,t^{\rm t}_{\rm sh}/t^{\rm t}_{\rm dif}),
\label{etasht}
\ee
where $t^{\rm t}_{\rm dif}$ is the diffusion time at the tail shock location and $t^{\rm t}_{\rm sh}$ is the duration of the shock.

Estimating the shock luminosities from equations \eqref{lshs} and \eqref{lsht} implicitly assumes that most of the radiation is released shortly after the self-crossing points, neglecting any emission close to the black hole. This assumption is legitimate for the following reasons. As will be demonstrated in Section \ref{cooling}, except in the ideal case where cooling is completely efficient, the stream rapidly expands under pressure forces shortly after its passage through the shock. This expansion induces a decrease of the thermal energy available for radiation as the stream leaves the shock location. Additionally, the radiative efficiency is likely lowered close to the black hole due to an shorter dynamical time, which reduces the emission in this region.

The radiative efficiencies $\eta^{\rm s}_{\rm sh}$ and $\eta^{\rm t}_{\rm sh}$ at the location of the self-crossing and tail shocks, given by equations \eqref{etashs} and \eqref{etasht} respectively, are a priori different since these two categories of shocks can happen at different positions. At early times, we nevertheless argue that they occur at similar locations. A justification can be seen in Fig. \ref{fig2} and \ref{fig6} where the blue solid line represents a parabolic trajectory with pericentre $\rp$, equal to that of the star. Because the debris in the tail are on elliptical orbits, their trajectories must be contained between this line and the orbit of the most bound debris, whose apocentre is indicated by a green star. The tail shocks therefore occur in the region delimited by these two trajectories. Since the self-crossing points (purple dots) are initially also located in this area, we conclude that the two radiative efficiencies are similar, with $\eta^{\rm s}_{\rm sh} \approx \eta^{\rm t}_{\rm sh}$ early in the stream evolution. At late times, when the stream orbit has precessed significantly, the self-crossing points leave this region possibly implying a significant difference between the two radiative efficiencies, with $\eta^{\rm s}_{\rm sh} \neq \eta^{\rm t}_{\rm sh}$. The time at which this happens is indicated by a vertical yellow dashed segment in Fig. \ref{fig8}. Another possibility is that the radiative efficiency decreases as the self-crossing points move closer to the black hole. In the following, we therefore estimate the effect of varying shock radiative efficiencies.

In the remainder of this section, values of $\eta^{\rm s}_{\rm sh}$ and $\eta^{\rm t}_{\rm sh}$ close to 1 are adopted, which corresponds to a case of efficient cooling where most of the thermal energy released by shocks is instantaneously radiated. Lower radiative efficiencies would imply lower shock luminosities. However, the shape of the total luminosity $L^{\rm s}_{\rm sh}+L^{\rm t}_{\rm sh}$ remains unchanged as long as $\eta^{\rm s}_{\rm sh} \approx \eta^{\rm t}_{\rm sh}$. If cooling is inefficient, a significant amount of thermal energy remains in the stream. The influence of this thermal energy excess on the subsequent stream evolution will be evaluated in Section \ref{cooling}.

The two other contributions to the luminosities given by equations \eqref{lshs} and \eqref{lsht} are orbital energy losses and mass rates through the shock. It is informative to examine the ratio between these quantities in the two shock luminosity components. The mass rate involved in the self-crossing shocks dominates that of tail shocks, with $\dot{M}_{\rm s} \gg\dot{M}_{\rm fb} $ typically after the first stream intersection. This tends to increase $L^{\rm s}_{\rm sh}$ compared to $L^{\rm t}_{\rm sh}$. Since the velocities involved in the two sources of shocks differ by at most $\sqrt{2}\approx 1.4$, the change of momentum experienced by the stream during tail shocks can be neglected compared to that imparted by self-crossing shocks. This justifies a posteriori our assumption of neglecting the dynamical influence of the tail on the stream evolution. The energy losses, on the other hand, are generally larger for the tail shocks, with $\epsilon_{\rm s} \gg \Delta \epsilon_{\rm s}$. This favours $L^{\rm t}_{\rm sh}$ larger than $L^{\rm s}_{\rm sh}$. It is therefore not obvious a priori which shock luminosity component dominates, which motivates the precise treatment presented below.

Fig. \ref{fig8} shows the temporal evolution of the stream shock luminosity $L^{\rm s}_{\rm sh}$ (red dashed line), tail shock luminosity $L^{\rm t}_{\rm sh}$ (blue dashed line), given by equations \eqref{lshs} and \eqref{lsht} respectively, and total shock luminosity $L^{\rm s}_{\rm sh}+L^{\rm t}_{\rm sh}$ (solid black line) for $\mh=10^5\msun$ (left panel) and $10^6\msun$ (right panel) assuming a flat energy distribution for the fallback rate and a stream dissipation timescale ${\Delta t}_{\rm dis} = P_{\rm s}$ in equation \eqref{mdots}. Equal radiative efficiencies are adopted for the two shock sources, with $\eta^{\rm s}_{\rm sh} = \eta^{\rm t}_{\rm sh}= 1$. The other two parameters are fixed to $\beta=1$ and $\vaovc=0$. For $\mh=10^5\msun$, the tail shock luminosity strongly dominates for $t/\tmin\lesssim20$. As this luminosity is proportional to the fallback rate $\dot{M}_{\rm fb} \propto t^{-5/3}$, the total shock luminosity also decreases as $t^{-5/3}$ following the solid orange segment. For $t/\tmin\gtrsim1$, the stream shock luminosity becomes dominant resulting in an increase of the total shock luminosity. For $\mh=10^6\msun$, the tail shock luminosity only weakly dominates initially, leading to a total shock luminosity only slightly decreasing for $t/\tmin\lesssim1$ before increasing at later times. The reason for $L^{\rm t}_{\rm sh}$ to drive the total shock luminosity at early times only for $\mh=10^5\msun$ relates to the stream dynamical evolution discussed in Section \ref{dynamical}. For lower black hole masses, the stream evolution is slower (see Fig. \ref{fig7}). The stream therefore retains a long period $P_{\rm s}$, which translates into a large value of the dissipation timescale ${\Delta t}_{\rm dis} = P_{\rm s}$. As a result, the stream mass rate $\dot{M}_{\rm s}$ through the shock diminishes (equation \eqref{mdots}) leading to a lower stream shock luminosity. On the other hand, the tail shock luminosity is unaffected since its temporal evolution is set by the fallback rate $\dot{M}_{\rm fb}$, independent of the stream evolution timescale. Fig. \ref{fig8} also shows the total luminosity for larger values of the penetration factor $\beta=2$ (solid purple line) and magnetic stresses efficiency $\vaovc=0.08$ (solid green line) keeping the other parameters fixed. Since this implies a faster stream evolution (see Fig. \ref{fig7}), the duration of the light curve decay is reduced for $\mh=10^5 \msun$ and even suppressed for $\mh=10^6 \msun$. A similar behaviour is noticed when the more precise fallback rate evolution of \citet{lodato2009} is adopted, assuming a polytopic star with $\gamma=5/3$ (grey solid line). For $\mh=10^5\msun$, the only difference is the presence of an initial increase towards a peak in the total luminosity evolution, reached at $t/\tmin\approx2$. This peak corresponds to the peak in the fallback rate, which the total luminosity follows initially. For $\mh=10^6\msun$, this more accurate fallback rate evolution results in a total luminosity always increasing since the fallback rate peaks at $t/\tmin \approx 2$, where the stream shock luminosity already dominates.

The shaded regions in Fig. \ref{fig8} show the areas covered by the total luminosity as $L^{\rm t}_{\rm sh}$ (red area) and $L^{\rm s}_{\rm sh}$ (blue area) are decreased up to a factor of 10 (red and blue solid lines). The former can occur if the radiative efficiencies satisfy $\eta^{\rm t}_{\rm sh} < \eta^{\rm s}_{\rm sh}$, which implies a decrease of the tail shock luminosity. The latter can be associated to $\eta^{\rm s}_{\rm sh} < \eta^{\rm t}_{\rm sh}$ or to a stream dissipation timescale such that ${\Delta t}_{\rm dis} > P_{\rm s}$, which both leads to a lower stream shock luminosity. Decreasing $L^{\rm t}_{\rm sh}$ leads to an initially lower total luminosity. Since the stream shock luminosity dominates earlier, the decay time is also quenched for $\mh=10^5 \msun$ and even removed for $\mh=10^6 \msun$. Instead, decreasing $L^{\rm s}_{\rm sh}$ leads to a lower total luminosity at late times with a longer initial decay time. A decrease of the radiative efficiency as self-crossing points get closer to the black hole would instead lead to a steeper decay of the total shock luminosity.

The total shock luminosity computed above remain mostly sub-Eddington. It is significantly lower than that $L_{\rm d}$ associated to the viscous accretion of a circular disc, assuming its rapid formation around to the black hole. For $\mh = 10^6 \msun$, the former peaks at $L_{\rm d}/L_{\rm Edd} \approx 100 $ while the latter only reaches $(L^{\rm t}_{\rm sh}+ L^{\rm s}_{\rm sh})/ L_{\rm Edd}\approx 0.1$. For this reason, the shock luminosity has been proposed by \citet{piran2015} as the source of emission from optical TDEs, detected at low luminosities. They also argue that this origin could explain the $t^{-5/3}$ decay of the optical light curve, as detected from this class of TDEs \citetext{e.g. \citealt{arcavi2014}}. However, this is only true if the tail shock luminosity component dominates. According to Fig. \ref{fig8}, this requires $\mh\lesssim 10^6 \msun$ for the most favourable values of the other two parameters, $\beta=1$ and $\vaovc=0$. In addition, this luminosity component is suppressed if magnetic stresses are efficient since they cause the stream to evolve faster. This makes the ballistic accretion scenario proposed by \citet{svirski2015} difficult to realize simultaneously with the decreasing optical luminosity.

\subsection{Impact of inefficient cooling}
\label{cooling}

\begin{figure}
\centering
\epsfig{width=0.4\textwidth, file=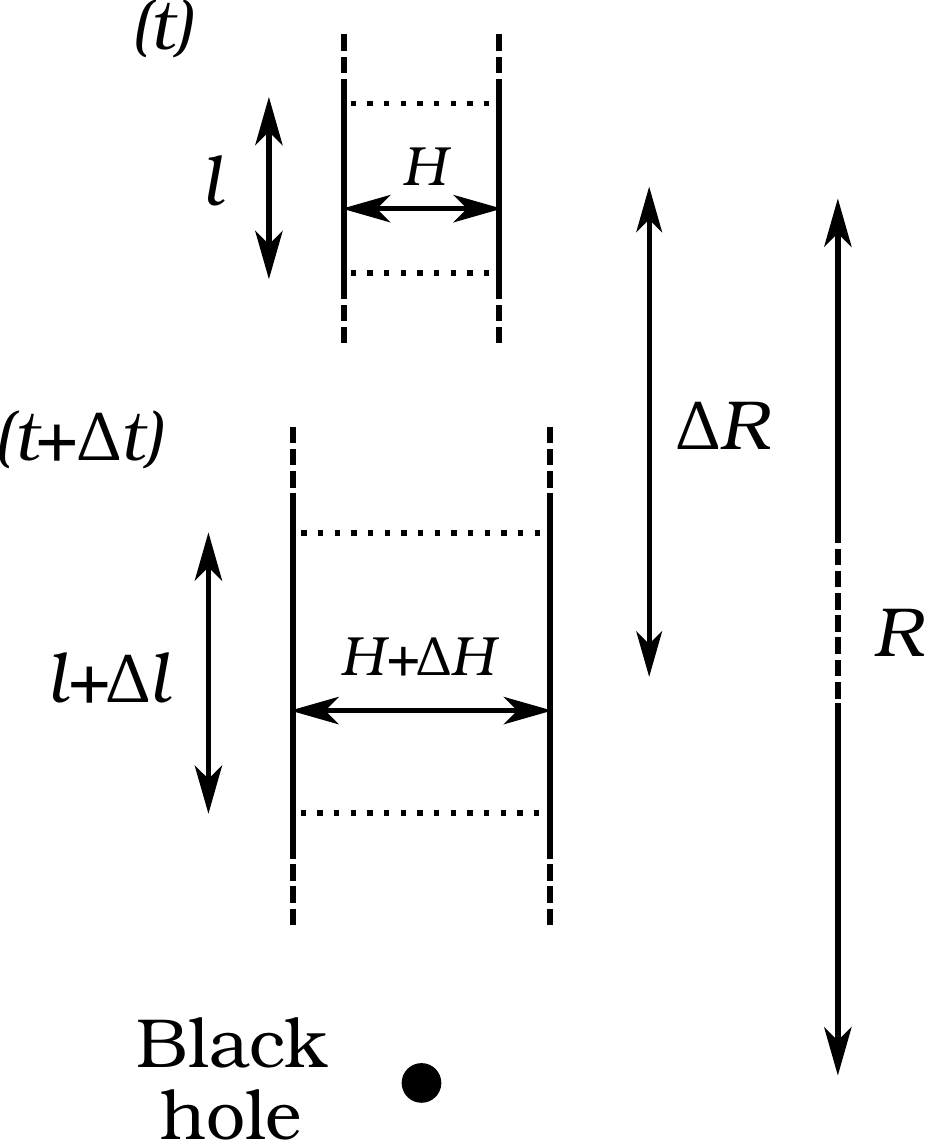}
\caption{Sketch illustrating the width evolution of a stream element as it approaches the black hole. At a time $t$, the element is located at a distance $R$ from the black hole with a width $H$ and a length $l$. After a time $\Delta t$, the element got closer to the black hole by a distance $\Delta R$ while its width and length varied by $\Delta H$ and $\Delta l$.}
\label{fig9}
\end{figure}

In Section \ref{observational}, values close to 1 have been adopted for the shock radiative efficiencies $\eta^{\rm s}_{\rm sh}$ and $\eta^{\rm t}_{\rm sh}$, artificially allowing most of the thermal energy injected by shocks to be released instantaneously in the form of radiation. Relaxing this assumption, part of the thermal energy stays in the stream, leading to its expansion through pressure forces. In this Section, we estimate this widening of the stream due to both stream self-crossing shocks and shocks between stream and tail. 

We start by deriving differential equations that relate the width of a stream element to its distance from the black hole. The situation is illustrated in Fig. \ref{fig9} in the case where the stream element is moving towards the black hole. At a given time $t$, the element has a width $H$ and is located at a distance $R$ from the black hole. A time $\Delta t$ later, its distance from the black hole decreased by $\Delta R<0$ while its width changed by $\Delta H$. Since the width evolves under the influence of both tidal and pressure forces, its variation can be decomposed into two components $\Delta H = \Delta H_{\rm t} + \Delta H_{\rm p}$, where $\Delta H_{\rm t}$ and $\Delta H_{\rm p}$ denote respectively the change of width due to tidal and pressure forces. As the stream element moves closer to the black hole, tidal forces induce a decrease of its width by $\Delta H_{\rm t} = - v_{\perp} \Delta t$ where $v_{\perp}$ is the velocity of the external part of the stream, directed towards the stream centre. For a nearly radial trajectory, $v_{\perp}$ can be related to the radial velocity $v_{\rm r}$ of the stream via $v_{\perp} \approx (H/R) v_{\rm r}$, which yields $\Delta H_{\rm t} = (H/R) \Delta R < 0$ using $\Delta R = - v_{\rm r} \Delta t$. Pressure forces cause the stream element to increase by $\Delta H_{\rm p} = c_{\rm s} \Delta t$, where $c_{\rm s}$ is the sound speed. Using $\Delta t = -\Delta R/v_{\rm r}$ then leads to $\Delta H_{\rm p}= -(c_{\rm s}/v_{\rm r}) \Delta R > 0$. Putting tidal and pressure components together, $\Delta H = (H/R - c_{\rm s}/v_{\rm r}) \Delta R$ if the stream element approaches the black hole. For a stream element moving away from the black hole, this relation becomes $\Delta H = (H/R + c_{\rm s}/v_{\rm r}) \Delta R$. The change of sign is required since $\Delta R>0$ in this case while pressure forces must still induce a increase of $H$. The evolution of the stream width $H$ as a function of distance $R$ from the black hole therefore obeys the differential equations 
\be
\frac{\diff H}{\diff R} = \cases{ \displaystyle \frac{H}{R}- \frac{c_{\rm s}}{v_{\rm r}} & \mbox{(inwards)}  \vspace{0.1cm}   \cr  \displaystyle\frac{H}{R}+ \frac{c_{\rm s}}{v_{\rm r}} & \mbox{(outwards)} \cr },
\label{dhovdr}
\ee
The first equation is valid when the stream moves inwards, towards the black hole. Instead, the second one corresponds to an outward motion of the stream, moving away from the black hole. In each equation, the first term on the right-hand side corresponds to the effect of tidal forces. Alone, it leads to an homologous evolution of the stream width, with $H\propto R$. This scaling can also be obtained by combining equations 4 and 13 of \citet{sari2010}. Instead, the second term is associated to pressure forces, which cause the stream expansion. Strictly speaking, our treatment of tidal effects is only valid for a nearly radial trajectory. However, our evaluation of pressure effects also applies to an elliptic orbit as long as $v_{\rm r}$ denotes the radial component of the total velocity. This method is legitimate since the stream trajectory significantly differs from a nearly radial one only at apocentre where pressure effects are found to be dominant.

For later use, we also derive the evolution of the stream element length $l$ with the distance $R$ from the black hole, still in the situation shown in Fig. \ref{fig9}. Pressure does not modify the element length since an expansion in the longitudinal direction is prevented by neighbouring stream elements. However, as the stream element moves closer to the black hole, its length increases by $\Delta l$ due to tidal forces. More precisely, this elongation is caused by the difference of velocity within the stream element, whose parts closer to the black hole move faster. The distance $|\Delta R|$ travelled during $\Delta t$ therefore becomes a function of $R$. The relative increase of length is then given by $\Delta l/l = - \diff |\Delta R|/\diff R$. The equation describing the element length evolution can then be written as a function of the radial velocity $v_{\rm r}$, such that $\diff l/l = \diff v_{\rm r}/v_{\rm r}$. $l$ therefore follows the same scaling as $v_{\rm r}$ with $R$, that is $l \propto R^{-1/2}$ for a nearly radial orbit. Again, this scaling can be found from \citet{sari2010}, combining their equations 4 and 14.

Our goal is now to solve equations \eqref{dhovdr} for boundary conditions defined by the dynamical model of Section \ref{model} that describes the stream evolution as a succession of orbits. As we show below, this allows to compute iteratively the width evolution of a stream element as it evolves around the black hole. Consider first a given orbit $N$. The stream element enters this orbit at apocentre, a distance $R^{\rm a}_N$ from the black hole with a velocity $v^{\rm a}_N$. It then approaches the black hole down to a pericentre distance $R^{\rm p}_N$ where its velocity reaches $v^{\rm p}_N$. These quantities are directly known from the dynamical model. In addition, the element has an initial width $H^{\rm a}_N$ and a sound speed $c^{\rm a}_{\rm s, \mathnormal{N}}$ at apocentre which can also be estimated as explained below. In order to solve equations \eqref{dhovdr}, the evolution of the second term $\pm c_{\rm s}/v_{\rm r}$ as the element follows orbit $N$ has to be evaluated. It requires to know the dependence on $H$ and $R$ of the sound speed $c_{\rm s,\mathnormal{N}}$ and radial velocity $v_{\rm r,\mathnormal{N}}$ in this orbit. These dependencies are obtained as follows. Assuming an adiabatic evolution, $c_{\rm s} \propto \rho^{1/3}$ with $\rho$ denoting the stream element density. Since the element keeps the same mass, the cylindrical profile of the stream imposes $\rho \propto H^{-2} l^{-1}$. The scaling $l\propto R^{-1/2}$ derived above for the element length then implies $c_{\rm s} \propto H^{-2/3} R^{1/6}$. The sound speed in orbit $N$ can therefore be written
\be
c_{\rm s,\mathnormal{N}} =  c^{\rm a}_{\rm s, \mathnormal{N}} \left(\frac{H}{H^{\rm a}_N}\right)^{-2/3} \left(\frac{R}{R^{\rm a}_N}\right)^{1/6}.
\label{csnev}
\ee
Similarly, the radial velocity of the element is obtained from
\be
v_{\rm r,\mathnormal{N}} =  \left(v^{\rm a}_N v^{\rm p}_N\right)^{1/2} \left( \frac{R^{\rm a}_N}{R}-1 \right)^{1/2} \left( 1-\frac{R^{\rm p}_N}{R} \right)^{1/2},
\label{vrnev}
\ee
which vanishes at apocentre and pericentre. The evolution of the stream element width $H$ with $R$ as it moves inwards from $R^{\rm a}_N$ to $R^{\rm p}_N$ is then obtained by solving numerically the first of equations \eqref{dhovdr}, using equations \eqref{csnev} and \eqref{vrnev} to evaluate the second term. The boundary condition is given by the element width $H^{\rm a}_N$ at $R=R^{\rm a}_N$. During the outwards motion of the stream away from pericentre to the next crossing point, the width evolution is found by solving the second of equations \eqref{dhovdr}, still combined with equations \eqref{csnev} and \eqref{vrnev}. In this phase, the boundary condition requires to know the width $H^{\rm p}_N$ at $R=R^{\rm p}_N$, which is obtained from the continuity of the element width at pericentre.

The last step to compute the element width evolution is to estimate the sound speed $c^{\rm a}_{\rm s,\mathnormal{N}}$ at apocentre that appears in equation \eqref{csnev}. Since shocks and magnetic stresses act at apocentre, both velocity and sound speed undergo discontinuous changes at this location. For the velocity, this discontinuity has already been accounted for in the dynamical model by computing $v^{\rm a}_N$ iteratively from its variation between two successive orbits $N$ and $N+1$, according to equations \eqref{vsh} and \eqref{vnew}. The sound speed $c^{\rm a}_{\rm s,\mathnormal{N}}$ can be evaluated in a similar iterative way. This requires to know the relation between the pre-shock sound speed $c^{\rm int}_{\rm s,\mathnormal{N}}$ of the stream element as it reaches the intersection point of orbit $N$ to the post-shock sound speed $c^{\rm a}_{\rm s,\mathnormal{N}+1}$ of the element as it leaves the shock location from the apocentre of orbit $N+1$. Since sound speed is related to specific thermal energy, this amounts to find the post-shock specific thermal energy $u^{\rm a}_{N+1}$ from its pre-shock value $u^{\rm int}_{N}$. This jump in thermal energy corresponds to the fraction of orbital energy lost through shocks from the tail and the stream that is not radiated away. Summing the two thermal energy components, this relation is
\be
u^{\rm a}_{N+1} = (1-\eta_{\rm sh})  \frac{\dot{M}^{\rm int}_{\rm s, \mathnormal{N}} (\Delta \epsilon_{\rm s, \mathnormal{N}}+u^{\rm int}_N) +  \dot{M}^{\rm int}_{\rm fb, \mathnormal{N}} \epsilon_{\rm s, \mathnormal{N}}}{\dot{M}^{\rm int}_{\rm s, \mathnormal{N}} + \dot{M}^{\rm int}_{\rm fb, \mathnormal{N}}},
\label{ush}
\ee
where $\dot{M}^{\rm int}_{\rm s, \mathnormal{N}}$ and $\dot{M}^{\rm int}_{\rm fb, \mathnormal{N}}$ are the mass rates at which the stream and the tail enters the shocks respectively, evaluated at the intersection point of orbit $N$. These factors account for the mass difference between the tail and stream components of the shock. $\Delta \epsilon_{\rm s, \mathnormal{N}}$ and $\epsilon_{\rm s, \mathnormal{N}}$ are the energies lost by the stream and the tail respectively during the shocks at orbit $N$, which are computed from the dynamical stream evolution. Equation \eqref{ush} also assumes that the stream self-crossing and tail shocks occur at the same position, leading to a common radiative efficiency $\eta_{\rm sh} \equiv \eta^{\rm s}_{\rm sh} = \eta^{\rm t}_{\rm sh}$ for the two shock sources. As explained in Section \ref{observational}, this approximation is legitimate, at least at early times. In general, the relation $u^{\rm int}_N \ll \Delta \epsilon_{\rm s, \mathnormal{N}}$ holds. The numerator of equation \eqref{ush} is therefore $(1-\eta_{\rm sh}) (L^{\rm s}_{\rm sh}+L^{\rm t}_{\rm sh}) / \eta_{\rm sh}$ (see equations \eqref{lshs} and \eqref{lsht}), which corresponds as expected to the thermal energy rate released by shocks but not radiated. Sound speed and specific thermal energy are linked via
\be
(c^{\rm a}_{\rm s,\mathnormal{N}})^2 = \frac{10}{9} u^{\rm a}_N,
\label{csa}
\ee
which allows to relate $c^{\rm a}_{\rm s,\mathnormal{N}+1}$ to $c^{\rm a}_{\rm s,\mathnormal{N}}$ using equation \eqref{ush}. $c^{\rm a}_{\rm s,\mathnormal{N}}$ can therefore be computed iteratively for any orbit $N$ starting from the pre-shock sound speed of the first shock, which we set to $c^{\rm int}_{\rm s,0} = 0$ since the stream element has not experienced any shock yet.

\begin{figure}
\epsfig{width=0.47\textwidth, file=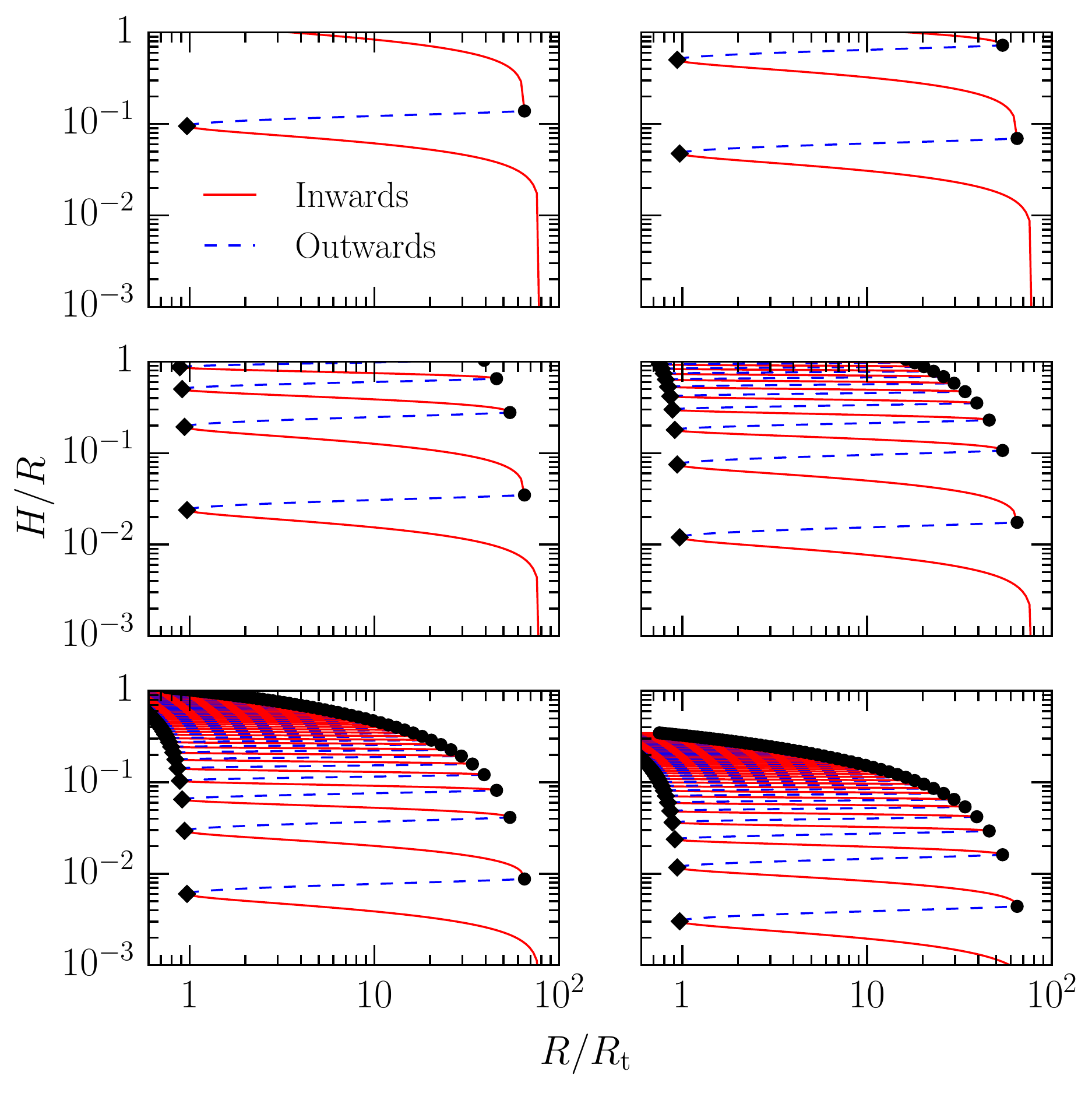}
\caption{Aspect ratio $H/R$ of a stream element as a function of distance from the black hole during the stream evolution. From the upper left to the lower right panel, the shock radiative efficiency increases from $\eta_{\rm sh} = 0$ to $1-\eta_{\rm sh} = 10^{-1}$, $10^{-2}$, $10^{-3}$, $10^{-4}$ and $10^{-5}$. The other parameters are fixed to $\mh=10^6 \msun$, $\beta=1$ and $\vaovc=0.06$. The corresponding  stream evolution is shown in Fig. \ref{fig2} (upper panel). The solid red lines correspond to an inward motion of the stream element from apocentre (black dot) to pericentre (black diamond), for which $H/R$ is obtained by solving the first of equations \eqref{dhovdr}. The dashed blue lines are associated to an outward motion of the stream, for which the second of equations \eqref{dhovdr} is solved to compute $H/R$.}
\label{fig10}
\end{figure}

\begin{figure}
\epsfig{width=0.47\textwidth, file=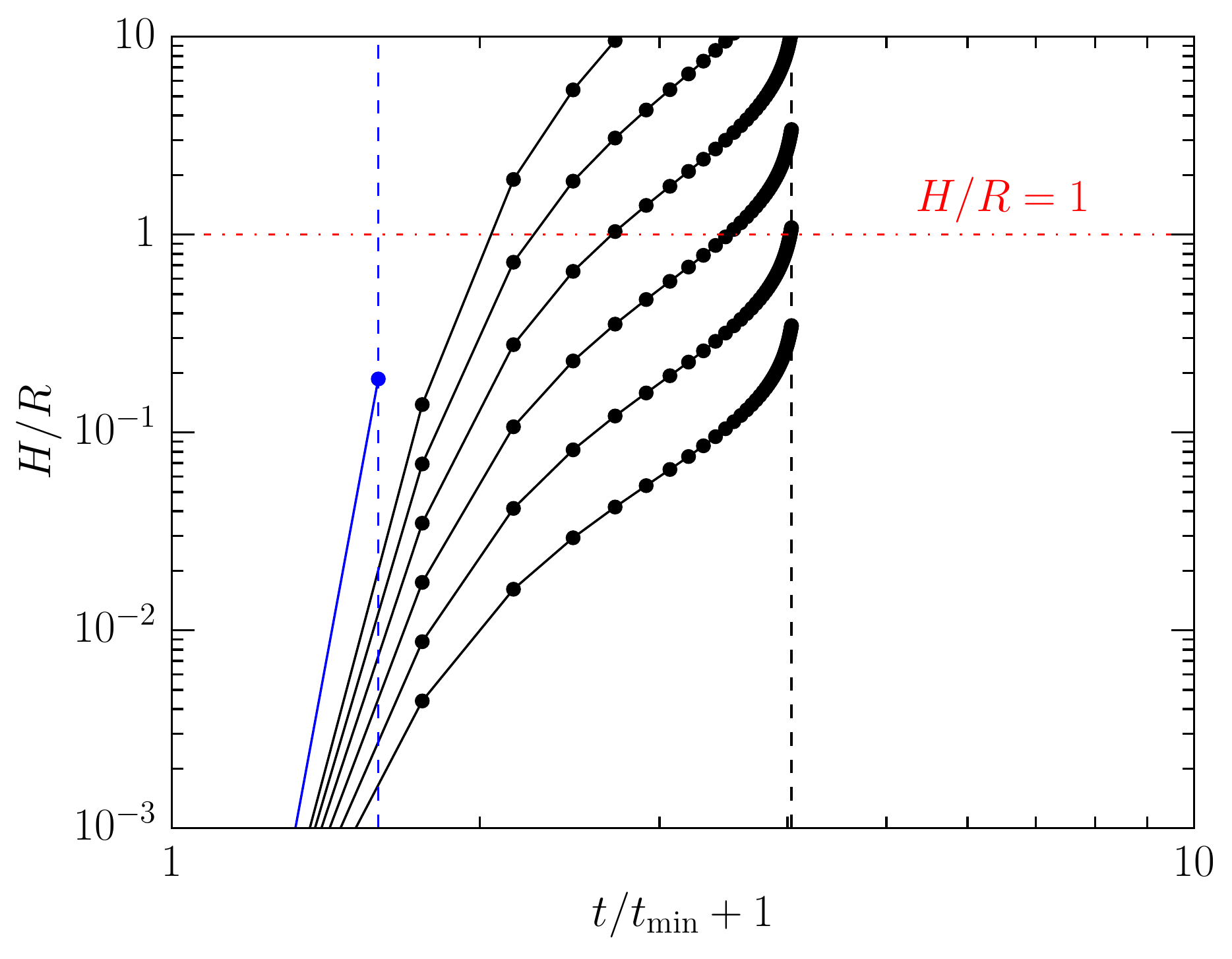}
\caption{Aspect ratio $H/R$ of a stream element as a function of time during the stream evolution. From the top to the bottom solid black line, the shock radiative efficiency $\eta_{\rm sh}$ increases from $\eta_{\rm sh} = 0$ to $1-\eta_{\rm sh} = 10^{-1}$, $10^{-2}$, $10^{-3}$, $10^{-4}$ and $10^{-5}$ (as in Fig. \ref{fig10}). The other parameters are fixed to $\mh=10^6 \msun$, $\beta=1$ and $\vaovc=0.06$. The blue line corresponds to a larger magnetic stresses efficiency $\vaovc=0.3$ for $\eta_{\rm sh} = 0$, keeping the other parameters fixed. The corresponding stream evolutions are shown in Fig. \ref{fig2}. The dots indicate apocentre passages. The vertical dashed lines represents the evolution time $t_{\rm ev}$ at which the stream ends its evolution in each case. The horizontal red dash-dotted line shows $H/R=1$.}
\label{fig11}
\end{figure}

The evolution of the width $H$ of a stream element as a function of $R$ during the stream evolution can now be computed iteratively assuming an initial value for $H$ and its continuity at self-crossing points. At the location of the first self-crossing point, a distance $R^{\rm int}_0$ from the black hole, the initial stream element width is fixed to $\rstar$. Although $H$ is imposed to be a continuous function of $R$, its derivative $\diff H/\diff R$ is discontinuous at apocentre and pericentre. This is because the differential equation satisfied by $\diff H/\diff R$ changes at these locations (see equation \eqref{dhovdr}). In addition, $c_{\rm s}$ increases instantaneously at apocentre where shocks occur and thermal energy is injected into the stream. Furthermore, since the radial velocity cancels at pericentre and apocentre, $\diff H/\diff R$ becomes infinite at these locations (see equations \eqref{dhovdr} and \eqref{vrnev}) resulting in a vertical tangent for $H$.

Our evaluation of the stream width evolution assumes that radiation only occurs near the self-crossing points and neglect any emission from the stream when it gets closer to the black hole. This allows to adopt an adabatic evolution for the stream element away from self-crossing points, which has been used to derive equation \eqref{csnev}. Note that this assumption has already been made in Section \ref{observational} to compute the shock luminosities. It is justified because the radiative efficiency likely decreases close to the black hole due to a shorter dynamical time in that region. If the stream is nevertheless able to cool at this location, pressure forces would be lowered reducing the stream expansion.

Fig. \ref{fig10} shows the aspect ratio $H/R$ of a stream element as a function of distance $R$ from the black hole during the stream evolution for increasing values of the shock radiative efficiency, from $\eta_{\rm sh} = 0$ (upper left panel) to $1-\eta_{\rm sh} = 10^{-1}$, $10^{-2}$, $10^{-3}$, $10^{-4}$ and $10^{-5}$ (lower right panel). The other parameters are fixed to $\mh=10^6 \msun$, $\beta=1$ and $\vaovc=0.06$. The corresponding stream evolution is shown in Fig. \ref{fig2} (upper panel). Initially, the stream element is located at the first self-crossing point, a distance $R^{\rm int}_0=80 \rt$ from the black hole (see Fig. \ref{fig2}, upper panel). The initial aspect ratio is $\rstar/R^{\rm int}_0 \approx 10^{-4}$ that is not visible on the figure. After the first shock, the aspect ratio increases and appears on the lower right part of each panel. It then continues to increase as the stream element successively moves inwards (solid red lines) and outwards (dashed blue line) between apocentre (black points) and pericentre (black diamonds). $H/R$ experiences a sharp increase shortly after each apocentre passage since thermal energy is injected into the stream at these locations (equation \eqref{ush}). Away from self-crossing points, the stream element width follows an almost homologous evolution $H \propto R$ since tidal forces dominate over pressure forces. As mentioned above, if the element was allowed to cool in this region, its width evolution would get even closer to an homologous one since pressure would be reduced. This would result in a slightly slower expansion of the stream. For $\eta_{\rm sh} = 0$ (upper left panel), the aspect ratio becomes $H/R>1$ after only a few (two) apocentre passages. For larger radiative efficiencies $\eta_{\rm sh}$, the aspect ratio increases more slowly. This is because less thermal energy is injected into the stream, which reduces the impact of pressure forces on the stream widening (equations \eqref{ush}, \eqref{csa} and \eqref{dhovdr}). The rapid aspect ratio increase seen for $\eta_{\rm sh} = 0$  persists as long as $1-\eta_{\rm sh} \gtrsim 10^{-3}$. As can be seen from Fig. \ref{fig11} (solid black lines), the aspect ratio reaches $H/R>1$ after a time $t \approx \tmin$, which corresponds to only a third of its evolution time $\tev/\tmin =3$. When the shock radiative efficiency satisfies $1-\eta_{\rm sh} \lesssim 10^{-4}$, the aspect ratio remains $H/R<1$ after a significant number of apocentre passages. Only for $1-\eta_{\rm sh} =10^{-5}$ (lower right panel), the stream circularizes with an aspect ratio $H/R<1$ (lower right panel of Fig. \ref{fig10} and bottom black solid line of Fig. \ref{fig11}).

Although the aspect ratio evolution is only shown for a particular set of parameters, this behaviour is similar for a large range of values for $\mh$, $\beta$ and $\vaovc$. A difference can nevertheless be noticed in the case of a rapid stream evolution, favoured for large values of these parameters (see Fig. \ref{fig7}). It can be understood by looking at the blue line of Fig. \ref{fig11}, for which the magnetic stresses efficiency is increased to $\vaovc=0.3$ compared to the top black line that shows $\vaovc=0.06$, both adopting a shock radiative efficiency $\eta_{\rm sh}=0$. For $\vaovc=0.3$, the stream experiences only two self-crossing before being ballistically accreted at $\tev/\tmin = 0.6$, as can be seen from the corresponding stream evolution shown in of Fig. \ref{fig2} (lower panel). As a result, only a small amount of thermal energy is injected in the stream, which results in an aspect ratio $H/R=0.6<1$ at the end of its evolution, even for $\eta_{\rm sh}=0$. This trend is also present for an increased black hole mass. For $\mh=10^7 \msun$, keeping $\beta=1$ and $\vaovc=0.06$, the stream circularizes with $H/R<1$ for lower radiative efficiencies than $\mh=10^6 \msun$, with a critical value $1-\eta_{\rm sh}  \approx 10^{-2}>10^{-5}$.

When the aspect ratio becomes $H/R \gtrsim 1$, pressure forces cannot anymore be neglected to describe the stream dynamics as we assume in our stream evolution model. This widening of the stream causes a large spread in its orbital parameters. The gas involved in the outward expansion moves to larger orbits while that expanding inwards gets shorter orbits. This is likely to cause complicated interactions between different portions of the stream, which are not captured by our model. The stream is likely to subsequently evolve into a thick torus, or even an envelope surrounding the black hole

\section{Discussion and conclusion}
\label{conclusion}

The dynamical evolution of the debris stream produced during TDEs is driven by two main mechanisms: magnetic stresses and shocks. Although these processes have been considered independently, their simultaneous effect has not been investigated. In this paper, we present a stream evolution model which takes both mechanisms into account. We demonstrate the existence of a critical magnetic stresses efficiency that sets the boundary between circularization and ballistic accretion. Interestingly, its value $(\vaovc)_{\rm cr} \approx 10^{-1}$ is found to be largely independent of the black hole mass and penetration factor. In the absence of magnetic stresses, we derive an analytical estimate for the circularization timescale $\tev/\tmin = 8.3 (\mh/10^6 \msun)^{-5/3} \beta^{-3}$. If magnetic stresses act on the stream, we prove that their dominant effect is to accelerate the stream evolution by strengthening self-crossing shocks. Ballistic accretion therefore necessarily occurs very early in the stream evolution. Instead, we show that a $t^{-5/3}$ decay of the shock luminosity light curve, likely associated to optical emission \citep{piran2015}, requires a slow stream evolution. This is favoured for low black hole masses $\mh \lesssim 10^6 \msun$ and hard to reconcile with the strong magnetic stresses necessary for the ballistic accretion scenario proposed by \citet{svirski2015}. Finally, we demonstrate that even marginally inefficient cooling with shock radiative efficiency $\eta_{\rm sh} \lesssim 1$ leads to the rapid formation of a very thick torus around the black hole, which could even evolve into an envelope encompassing it as proposed by several authors \citep{guillochon2014,metzger2015}. This thick structure could act as a reprocessing layer that intercepts a fraction of the X-ray photons released as debris accretes onto the black hole and re-emit them as optical light. In this picture, the detection of X-ray emission would however be dependent on the viewing angle. For example, the X-ray photons could still be able to escape along the funnels of a thick torus but not along its orbital plane.

In the absence of magnetic stresses, the stream evolution predicted by our model is qualitatively consistent with existent numerical simulations of disc formation from TDEs. Specifically, our model can be compared to \citet{bonnerot2016_2} who consider the disruption of bound stars by a non-rotating black hole of mass $\mh=10^6 \msun$. For a stellar eccentricity $e=0.95$ and penetration factor $\beta =5$, they find that disc formation occurs in less than a dynamical time due to strong apsidal precession. For $\beta =1$, their simulations predict instead a series of self-crossing shocks that leads to a longer circularization from an initially eccentric disc due to weaker apsidal precession. These two numerical results are in line with our analytic model, which considers the most general case of a parabolic stellar orbit. Finally, the rapid thickening of the stream predicted analytically in this paper in the case of inefficient cooling is consistent with several disc formation simulations that assume an adiabatic equation of state for the debris \citep{guillochon2014,shiokawa2015,bonnerot2016_2,hayasaki2015,sadowski2015}. However, these simulations have been performed in the restricted case of either very low black hole masses or bound stars for numerical reasons. Instead, our analytic model treats the standard case.

The range of magnetic stresses efficiencies $\vaovc$ investigated has been set by assuming saturation of the MRI. Although this assumption is legitimate in well-ordered discs after a few dynamical timescales, it is unclear whether it holds in the case of the debris stream that is likely to lose its ordering through shocks. If the MRI has not reached saturation, $\vaovc$ would be lowered thus reducing the dynamical effect of magnetic stresses and preventing ballistic accretion. If it goes down to $\vaovc \lesssim 10^{-2}$, magnetic stresses would even become dynamically irrelevant.

To estimate the widening of the stream, we investigate a wide range of radiative shock efficiencies. The values expected physically for this parameter can be determined by the following calculation. Assuming that the photons propagate in the direction transverse to the stream trajectory, the diffusion time for them to escape the stream is $t_{\rm dif} = c/(\tau H_{\rm sh})$ with $\tau$ and $H_{\rm sh}$ the optical depth and width at the shock location. The optical depth can be estimated by $\tau = \Sigma \kappa_{\rm T}$ where $\Sigma$ denotes the column density at the shock location and $\kappa_{\rm T}=0.4 \, \rm cm^2/g$ is the Thomson opacity. The column density can be approximated by $\Sigma \approx \dot{M} t_{\rm sh} /  H^2_{\rm sh}$, denoting by $\dot{M}=\dot{M}_{\rm d}+\dot{M}_{\rm fb}$ the total mass rate of matter through the shock location. Combining these expressions, the shock radiative efficiency takes the form $\eta_{\rm sh} \equiv t_{\rm sh}/ t_{\rm dif} \approx H_{\rm sh} c/(\kappa_{\rm T}\dot{M})$. For the first self-crossing shock, $H_{\rm sh}\approx \rstar$ and $\dot{M} \approx \mstar/(3\tmin)$, which corresponds to the peak fallback rate. This leads to an initial radiative efficiency of $\eta_{\rm sh} \approx 10^{-5} (\mh/10^6\msun)^{1/2} (\rstar/R_{\odot})^{5/2} (\mstar/M_{\odot})^{-2}$, where the parameters adopted assume the disruption of a solar-type star. This implies that a very inefficient cooling is expected in this case. This estimate is consistent with recent radiative transfer simulations by \citet{jiang2016} focusing on the stream self-intersection region. They obtain values of the shock radiative efficiency of $\eta_{\rm sh}\approx 0.01-0.1$ for mass accretion rates roughly three orders of magnitude lower than used in our estimation. This expression also demonstrates that the shock radiative efficiency is expected to increase for disruptions involving red giants, up to $\eta_{\rm sh}\approx 1$ if the stellar radius reaches $\rstar \approx 100 \rsun$. For the following shocks, the dependence $\eta_{\rm sh} \propto H_{\rm sh} \dot{M}^{-1}$ implies the following evolution of the shock radiative efficiency. During the initial expansion of the stream, $H_{\rm sh}$ rapidly increases (see Fig. \ref{fig11}) and so does $\eta_{\rm sh}$, most likely by a few orders of magnitude. Later in time, the stream expansion stalls. The increase of $\dot{M} \approx \dot{M}_{\rm d}$ then dominates and eventually induces a drop of $\eta_{\rm sh}$. This temporal dependence of $\eta_{\rm sh}$ is unlikely to significantly affect our results for the width evolution of the stream. However, it induces a modulation of the shock luminosity since the amount of radiation able to diffuse out of the stream depends on the shock radiative efficiency.

The only effect of magnetic stresses considered in our model is angular momentum loss at apocentre, thus increasing further the stream eccentricity. However, another possibility has been pointed out by numerical simulations. It was found that small-scale instability can develop in eccentric discs that damps the eccentricity \citep{papaloizou2005,barker2016}.

Finally, we neglect for simplicity the black hole spin in our calculations. If the orbital plane of the debris is not orthogonal to the black hole spin, it causes the stream to  change orbital plane when it passes at pericentre. One possible consequence is to delay the onset of circularization by preventing the first self-crossing \citep{dai2013,guillochon2015}. However, it can also lead to faster energy dissipation due to complicated interactions between parts of the stream belonging to different orbital planes \citep{hayasaki2015}.

Our model provides a first attempt at studying the evolution of the debris stream under both shocks and magnetic stresses. It is attractive by its simplicity and points out several solid features about the dynamics, observational appearance and geometry of the stream as it evolves around the black hole. However, given the complexity of this process and the numerous physical mechanisms involved, global simulations are necessary to definitely settle the fate of these debris.


\label{lastpage}

\bibliography{biblio}

\section*{Appendix A: Analytic evolution time}

Here, we derive the analytical expression for the evolution time in the absence of magnetic stresses, given by equation \eqref{tevana}. The evolution time is defined by the time spent in all the orbits followed by the stream during the succession of ellipses described in Section \ref{model}. Mathematically,
\be
\tev = \sum_{N=0}^{N_{\rm ev}} P_N,
\label{tevdef}
\ee
where $N_{\rm ev}$ is the total number of orbits followed by the stream and $P_N$ is the period of the stream in orbit $N$. Since period and energy are related by $P = 2 \pi G \mh (-2 \epsilon)^{-3/2}$ according to Kepler's third law, $\diff P^{1/3}/\diff \epsilon =  (2 \pi G \mh)^{-2/3} P$ whose discretized version can be written
\be
\frac{P^{1/3}_{N+1} - P^{1/3}_N}{\Delta \epsilon_N} = -(2 \pi G \mh)^{-2/3} P_N,
\label{recurseq}
\ee
where $\Delta \epsilon_N$ is the energy lost by the stream at each shock. As explained at the end of Section \ref{shocks}, $\Delta \epsilon_N$ is independent of $N$ if no magnetic stresses act on the stream. Its value is then $\Delta \epsilon_0$, given by equation \eqref{deltaenzero}. Using $\tmin = 2 \pi G \mh (2 \Delta \epsilon)^{-3/2}$, equation \eqref{recurseq} can therefore be rewritten
\be
P^{1/3}_{N+1} - P^{1/3}_N =  -\frac{ \Delta \epsilon_0}{2 \Delta \epsilon} \frac{P_N}{t^{2/3}_{\rm min}}.
\ee
Summing both sides from $N=0$ to $N_{\rm ev}$, the terms on the left-hand side cancel two by two leading to
\be
P^{1/3}_{N_{\rm ev} +1} - P^{1/3}_0 = -\frac{ \Delta \epsilon_0}{2 \Delta \epsilon} \frac{\tev}{t^{2/3}_{\rm min}},
\label{summedeq}
\ee
where equation \eqref{tevdef} has been used to write $\tev$ on the right-hand side. As demonstrated in Section \ref{results}, the final outcome of the stream is circularization for $\vaovc = 0$ implying $P^{1/3}_{N_{\rm ev} +1} \ll P^{1/3}_0$. Therefore, equation \eqref{summedeq} can be simplified to
\be
\frac{\tev}{\tmin} = \frac{2 \Delta \epsilon}{ \Delta \epsilon_0},
\ee
using the fact that the initial stream period is $P_0 = \tmin$. This demonstrates the analytical expression for the evolution time.

\end{document}